\documentclass[prd]{revtex4}
\usepackage{float}
\usepackage{graphicx}
\usepackage{caption}
\usepackage{amsmath}
\usepackage{array}
\usepackage{bm}
\usepackage{amssymb}
\usepackage{amsfonts}
\usepackage{amssymb}
\usepackage{color}

\newcommand{\mP}{\mathcal{P}}
\newcommand{\mV}{\mathcal{V}}
\newcommand{\mA}{\mathcal{A}}
\newcommand{\mS}{\mathcal{S}}

\begin{document}
\title{Spin evolution of massive fermion in QED plasma}
\author{Ziyue Wang}
\affiliation{Department of Physics, Tsinghua University, Beijing 100084, China}
\email{zy-wa14@mails.tsinghua.edu.cn}
\date{\today}
\begin{abstract}
The dynamical evolution of spin of a massive probe fermion in a massless hot QED plasma at local equilibrium is investigated through the quantum kinetic theory. We consider the massive probe fermion undergoing 2-by-2 Coulomb scattering with the massless fermions in the medium. The axial kinetic equation is derived including the collision terms to the first order of gradients and leading logarithmic order of the coupling. The collision terms are vanishing at global equilibrium, around which the relaxation time can be extracted as an operator. We further decompose the axial kinetic equation into kinetic equations of axial-charge density as well as the transverse magnetic dipole moment, and illustrate the diffusion and polarization effect through preliminary numerical analysis.
\end{abstract}
\maketitle
\section{Introduction}
\label {s1}
Recent STAR and ALICE experiments \cite{STAR:2017ckg, STAR:2019erd, ALICE:2019aid, Singha:2020qns, STAR:2020xbm} have shed light on the spin polarization of hadrons in the rotating QCD plasma produced in off-central relativistic heavy-ion collisions. Such spin polarization of emitted hadrons \cite{Liang:2004ph, Liang:2004xn, Gao:2007bc, Becattini:2007sr} has motivated researches concerning the dynamical evolution of spin for particles in a finite temperature plasma. Part of the large initial orbital angular momentum characterized by the collective motion of the fluid is transferred to the spin of the particles through collisions. The polarized particles also experience relaxation processes that drive the spin polarization to equilibrium. The global polarization of $\Lambda$ hyperons enslaved by the thermal vorticity \cite{Becattini:2013fla} is a robust phenomenon, where model calculations \cite{Li:2017slc, Wei:2018zfb, Karpenko:2016jyx, Xie:2017upb, Sun:2017xhx, Ryu:2021lnx} are in consistency with experiments. However, such satisfaction has not been achieved in local spin polarization. The measurement of azimuthal angle dependence of spin polarization in experiments \cite{STAR:2019erd} has not been fully understood in theoretical studies due to the opposite sign in the phenomenological studies assuming the global equilibrium of spin\cite{Becattini:2017gcx, Xia:2018tes}. Such inconsistency is also known as the spin sign problem. Attempts to resolve this problem include modifying the understanding of vorticity \cite{Wu:2019eyi}, feed-down effect \cite{Li:2021zwq, Xia:2019fjf}, hyperon decay \cite{Becattini:2019ntv}. It is realized later the inclusion of shear tensor in the polarization yields the qualitatively correct sign \cite{Fu:2021pok, Becattini:2021iol, Liu:2021uhn, Becattini:2021suc}, indicating off-equilibrium effects of spin maybe essential in polarization phenomenon. It is also found that the numerical results could be sensitive to the parameters in numerical analysis\cite{Yi:2021ryh, Florkowski:2021xvy, Sun:2021nsg, Wu:2022mkr}. This calls for more thorough investigations of the non-equilibrium effects in the spin polarization. 

Theoretical description of the dynamical evolution of spin polarization is mainly based on quantum kinetic theory \cite{Gao:2019znl, Weickgenannt:2019dks, Hattori:2019ahi, Wang:2019moi, Liu:2020flb, Li:2019qkf, Yang:2020hri, Weickgenannt:2020aaf, Wang:2020pej, Wang:2021qnt, Lin:2021mvw, Fang:2022ttm, Hidaka:2022dmn} and spin hydrodynamics \cite{Florkowski:2017ruc, Florkowski:2018fap, Hattori:2019lfp, Fukushima:2020ucl, Bhadury:2020puc, Shi:2020htn, Li:2020eon, Gallegos:2021bzp, She:2021lhe, Hongo:2021ona}. The chiral kinetic theory (CKT) \cite{Son:2012bg, Son:2012wh, Stephanov:2012ki, Son:2012zy, Chen:2013iga, Chen:2014cla, Chen:2015gta, Hidaka:2016yjf, Hidaka:2017auj, Huang:2018wdl, Liu:2018xip} was developed to describe the spin related anomalous transport phenomena, and has been applied to chiral magnetic effect\cite{Kharzeev:2004ey} in heavy ion collisions. It is then extended to the quantum kinetic theory to describe the spin transport of massive fermions \cite{Gao:2019znl, Weickgenannt:2019dks, Hattori:2019ahi, Wang:2019moi}. In recent years, the collision terms are also included to study the relaxation process of spin \cite{Yang:2020hri, Weickgenannt:2020aaf, Wang:2020pej, Wang:2021qnt, Lin:2021mvw, Fang:2022ttm}.  The general framework of quantum kinetic theory is based on the Wigner function and Keldysh formalism, which is able to keep the full power of quantum field theory in non-equilibrium system \cite{Blaizot:2001nr}. On the other hand, spin hydrodynamics extends the standard conservation laws to also include the conservation of angular momentum, describes the macroscopic evolution of spin density. 

The polarization of $\Lambda$ hyperons is dominated by the s-quark, which can not be approximated as massless fermion. In order to investigate the spin dynamics of s-quark in the quark gluon plasma, we in this work deal with a simplified scenario as a first step to the full problem. We consider the evolution of spin of a hard massive fermion $m\gg eT$ probing into a hot massless QED plasma at local equilibrium. As the Compton scattering is suppressed in case $m\gg eT$, and the evolution is dominated by Coulomb scattering. Two competing processes would contribute to the spin evolution, the diffusion process coming from the scattering with medium fermion drives the fluctuation of spin back to equilibrium, while the collective motion of the medium, characterized by the hydrodynamic gradients, acts as a source to polarize the spin of massive fermion. So as to self-consistently incorporate the two processes, we derive the collision terms to $\mathcal{O}(\hbar)$ with all the first order hydrodynamic gradients included. In \cite{Li:2019qkf, Yang:2020hri, Hongo:2022izs}, only classical processes characterizing the diffusion processes are discussed; in \cite{Wang:2021qnt} the polarization effect of vorticity is discussed while the contact interaction of NJL model is not enough to catch the dynamical process in QGP; in \cite{Fang:2022ttm} the collision term is derived to $\mathcal{O}(\hbar)$, while the probe fermion is massless. In this paper, we derive the collision term to the leading logarithmic order in coupling $e$, similar to the procedure in \cite{Li:2019qkf, Yang:2020hri, Hongo:2022izs, Fang:2022ttm}. For the massive fermion, the axial-vector component of Wigner function characterizing the spin distribution has three degrees of freedom. To pave the way for numerical calculation, we decompose the axial kinetic equation to kinetic equations of the axial-vector charge and the transverse dipole moment. Some preliminary numerical analysis is also presented to illustrate the diffusion and polarization processes described by the collision terms. 

This paper is organized as follows: in Sec.\ref{s2}, we briefly review the Wigner function and Kadanoff-Baym equations, as well as the power counting scheme. In Sec.\ref{s3}, we derive the general expression for the collision term and discuss contribution from the various part of the collision term. In Sec.\ref{s4}, the result of collision term after integral over phase space momentum is presented, together with expression in massless and non-relativistic limit; The relaxation rate near the global equilibrium is also extracted. In Sec.\ref{s5}, the axial kinetic equation is further decomposed into kinetic equation of axial-charge density and transverse dipole moment. A preliminary numerical analysis decorating the diffusion and polarization processes is presented. In Sec.\ref{s6}, we provide conclusion and outlook. Calculation details are presented in Appendix.\ref{calculation_detail} and \ref{gaugeissue}.

In this paper, we take the mostly-negative convention of metrix $g_{\mu\nu}=\text{diag}(1,-1,-1,-1)$ and take the Dirac matrix in the Weyl basis with $\gamma_5=i\gamma^0\gamma^1\gamma^2\gamma^3$ and $\sigma^{\mu\nu}=i[\gamma^\mu,\gamma^\nu]/2$. The Levi-Civita symbol is chosen as $\epsilon^{0123}=-\epsilon_{0123}=+1$. We use majuscule letter for four-dimension covariant momentum such as $P^\mu$ and use minuscule latter for its component such as $p^0$ and its module such as $p=|\vec{p}|$. We use the projector $\Delta^{\mu\nu}=g^{\mu\nu}-u^\mu u^\nu$ to project a vector onto direction perpendicular to the fluid velocity $u^\mu$, such as $P_\perp^\mu=\Delta^{\mu\nu}P_\nu$ and define $\hat{P}_\perp^\mu=P_\perp^\mu/p$ with $p=(-P_\perp^\mu P_{\perp\mu})^{1/2}$. The projector $\Xi^{\mu\nu}=g^{\mu\nu}-u^\mu u^\nu+\hat{P}_\perp^\mu\hat{P}_\perp^\nu$ projects a vector onto direction perpendicular to both $u^\mu$ and $P_\perp^\mu$. We also use the following notations for the first order gradients: $\theta=\partial\cdot u$, $D=u\cdot\partial$, the fluid vorticity defined as $\omega^\mu\equiv\frac{1}{2}\epsilon^{\mu\nu\alpha\beta}u_\nu\partial_\alpha u_\beta$ and shear tensor $\sigma^{\langle\alpha\beta\rangle}$ defined as the symmetric and traceless part of $\sigma^{\alpha\beta}=\frac{1}{2}(\partial_\perp^\alpha u^\beta+\partial_\perp^\beta u^\alpha)-\frac{1}{3}\Delta^{\alpha\beta}\theta$. The symmetrization and anti-symmetrization of two symbols are defined through $X_{(\alpha}Y_{\beta)}=X_{\alpha}Y_{\beta}+X_{\beta}Y_{\alpha}$ and $X_{[\alpha}Y_{\beta]}=X_{\alpha}Y_{\beta}-X_{\beta}Y_{\alpha}$.
\section{Spin transport equation}
\label{s2}
In this section, we review the basic steps of deriving the axial kinetic equation with collision term. Starting from the Wigner transformation applied to contour Green's function \cite{Blaizot:2001nr}
\begin{eqnarray}
S_{\alpha\beta}^{<(>)}(X,p)=\int d^4 Y e^{ip\cdot Y/\hbar}\tilde{S}_{\alpha\beta}^{<(>)}(x,y),
\end{eqnarray}
where $X=(x+y)/2$ and $Y=x-y$ are the center of mass coordinate and relative coordinate. Here, $\tilde{S}_{\alpha\beta}^{<}(x,y)=\langle\bar{\psi}_\beta(y)\psi_\alpha(x)\rangle$ and $\tilde{S}_{\alpha\beta}^{>}(x,y)=\langle\psi_\alpha(x)\bar{\psi}_\beta(y)\rangle$ are lessor and greater propagators, respectively. After Wigner transformation, the lessor propagator obeys the Kadanoff-Baym equations derived from the Schwinger-Dyson equation,  
\begin{eqnarray}
\label{KBeq}
\Big(\gamma^\mu P_\mu-m\Big)S^<
+\frac{i\hbar}{2}\gamma^\mu\nabla_\mu S^<
~=~\frac{i\hbar}{2}\Big(\Sigma^<\star S^>-\Sigma^>\star S^<\Big),
\end{eqnarray}
where $\Sigma^{>(<)}$ represents the lessor(greater) self-energy. The scattering process involves only $\Sigma^{<(>)}$, thus we have dropped the real parts of the retarded and advanced self-energies and of the retarded propagators. The electro-magnetic fields decay quickly in the QGP, hence we neglect the background electro-magnetic fields in the medium. The symbol $\star$ represents $A\star B=AB+\frac{i\hbar}{2}[AB]_{\text{P.B.}}+\mathcal{O}(\hbar^2)$, where the Poisson bracket is $[AB]_{\text{P.B.}}\equiv(\partial_q^\mu A)(\partial_\mu B)-(\partial_\mu A)(\partial_q^\mu B)$. The commutators are defined as $\{F,G\}\equiv FG+GF$, $[F,G]\equiv FG-GF$, $\{F,G\}_\star\equiv F\star G+G\star F$ and $[F,G]_\star\equiv F\star G-G\star F$ with $F$ and $G$ are arbitrary matrix-valued functions. By using the complete basis for the Clifford algebra, the Wigner function is decomposed into $S^<=\mathcal{S}+i\mathcal{P}\gamma^5+\mathcal{V}_\mu\gamma^\mu+\mathcal{A}_\mu\gamma^5\gamma^\mu+\frac{1}{2}\mathcal{S}_{\mu\nu}\sigma^{\mu\nu}$ and $S^>=\bar{\mathcal{S}}+i\bar{\mathcal{P}}\gamma^5+\bar{\mathcal{V}}_\mu\gamma^\mu+\bar{\mathcal{A}}_\mu\gamma^5\gamma^\mu+\frac{1}{2}\bar{\mathcal{S}}_{\mu\nu}\sigma^{\mu\nu}$. Similarly, it is also useful to carry out the same spinor-basis decomposition for the self-energies, giving $\Sigma^<=\Sigma_S+i\Sigma_P\gamma^5+\Sigma_{V\mu}\gamma^\mu+\Sigma_{A\mu}\gamma^5\gamma^\mu+\frac{1}{2}\Sigma_{T\mu\nu}\sigma^{\mu\nu}$ and $\Sigma^>=\bar{\Sigma}_S+i\bar{\Sigma}_P\gamma^5+\bar{\Sigma}_{V\mu}\gamma^\mu+\bar{\Sigma}_{A\mu}\gamma^5\gamma^\mu+\frac{1}{2}\bar{\Sigma}_{T\mu\nu}\sigma^{\mu\nu}$. $\mathcal{V}$ and $\mathcal{A}$ give rise to the vector-charge and axial-charge currents through $J_V^\mu=\int q\mathcal{V}^\mu$ and $J_5^\mu=\int q\mathcal{A}^\mu$. The axial-charge currents can be regarded as a spin current of fermion. Taking $\mathcal{V}$ and $\mathcal{A}$ as independent degrees of freedom, the scalar component $\mathcal{S}$, pseudo-scalar component $\mathcal{P}$ and tensor component $\mathcal{S}_{\mu\nu}$ can be expressed in terms of $\mathcal{V}$ and $\mathcal{A}$.

We are going to investigate the relaxation of spin of a massive probe fermion injecting into a hot massless QED plasma in local equilibrium. Before moving on to calculate the collision term, we first introduce the $\hbar$-counting, which is equivalent to counting in gradients. In the heavy ion collision, the axial-vector currents are mostly induced by the electro-magnetic field or the gradients of the fluid velocity. This motivates the counting of $\mathcal{A}_\mu\sim O(\hbar)$. On the other hand, the vector charge current can be safely kept only to $\mathcal{O}(\hbar^0)$, as it is dominated by classical process. The power-counting of $\mathcal{A}_\mu$ and $\mathcal{V}_\mu$ also leads to counting of the other components $\mathcal{S}\sim\mathcal{O}(\hbar^0)$, $\mathcal{S}_{\mu\nu}\sim\mathcal{O}(\hbar^1)$ and $\mathcal{P}\sim\mathcal{O}(\hbar^2)$, as well as the components of the self-energy. In the Coulomb scattering we are going to investigate, the above counting leads to $\Sigma_S\sim\mathcal{O}(\hbar^0)$, $\Sigma_{V\mu}\sim\mathcal{O}(\hbar^0)$, $\Sigma_{A\mu}\sim\mathcal{O}(\hbar^1)$, $\Sigma_{T\mu\nu}\sim\mathcal{O}(\hbar^1)$ and $\Sigma_P\sim\mathcal{O}(\hbar^2)$. The thermalization of the vector charge is dominated by the classical process, thus is enough to keep only $\mathcal{O}(\hbar^0)$ terms in the collision term. The thermalization of spin involves diffusion of the initial spin of the probe as well as polarization induced by gradients such as the vorticity and shear, it is required to evaluate the collision terms up to $\mathcal{O}(\hbar)$. Then the collision terms for vector and axial-vector components are be obtained though comparing the Dirac structures on both sides of Kadanoff-Baym equation, giving the vector kinetic equation
\begin{eqnarray}
\partial_\mu\mathcal{V}^{\mu}&=&
-\frac{P_\mu}{m}\widehat{\Sigma_S\mathcal{V}^\mu}
-\widehat{\Sigma_{V\mu}\mathcal{V}^{\mu}}+O(\hbar),
\end{eqnarray}
where $\widehat{XY}=\bar{X}Y-X\bar{Y}$. And the axial kinetic equation
\begin{eqnarray}
\label{Atransport}
P\cdot\partial\mathcal{A}_\mu&=&
-m\widehat{\Sigma_S\mathcal{A}_{\mu}}
-P^\nu\widehat{\Sigma_{V\nu}\mathcal{A}_{\mu}}
-P^\nu\widehat{\Sigma_{A\mu}\mathcal{V}_\nu}
-\frac{m}{2}\epsilon_{\alpha\beta\lambda\mu}\widehat{\Sigma_T^{\alpha\beta}\mathcal{V}^{\lambda}}
+P_\mu\widehat{\Sigma_{A\nu}\mathcal{V}^{\nu}}
+\frac{1}{2}\epsilon_{\mu\nu\rho\sigma}(\partial^\sigma\widehat{\Sigma_{V}^{\nu})\mathcal{V}^{\rho}}+O(\hbar^2).
\end{eqnarray}
The power counting $\mathcal{A}_\mu\sim\mathcal{O}(\hbar)$ guarantees the mass-shell condition of $\mathcal{A}_\mu$ \cite{Wang:2020pej}, see also \cite{Yang:2020hri} for details of derivation. With the relations between various components of the Wigner function, the parameterization of the various components can be taken as,
\begin{eqnarray}
\label{Wignercompo}
\mathcal{S}&=&2\pi\epsilon(P\cdot u)\delta(P^2-m^2)mf_V,\nonumber\\
\mathcal{V}_\mu&=&2\pi\epsilon(P\cdot u)\delta(P^2-m^2)P_\mu f_V,\nonumber\\
\mathcal{A}_\mu&=&2\pi\epsilon(P\cdot u)\delta(P^2-m^2)n_\mu,\nonumber\\
\mathcal{S}_{\mu\nu}&=&2\pi\epsilon(P\cdot u)\delta(P^2-m^2)\Sigma_{\mu\nu}.
\end{eqnarray}
We do not take any decomposition of $n_\mu$ at the moment, for now it is only constrained by $P^\mu n_\mu=0$ coming directly from $P^\mu \mathcal{A}_\mu=0$. With the relation $\mathcal{S}_{\mu\nu}=\frac{1}{2m}\partial_{[\mu}\mathcal{V}_{\nu]}-\frac{1}{m}\epsilon_{\mu\nu\rho\sigma}P^\rho\mathcal{A}^{\sigma}+\mathcal{O}(\hbar^2)$ between the tensor and axial-vector component \cite{Wang:2020pej}, $\Sigma_{\mu\nu}$ is expressed as 
\begin{eqnarray}
\Sigma_{\mu\nu}(P)&=&-\frac{1}{2m}P_{[\mu}\partial_{\nu]}f_V(P)-\frac{1}{m}\epsilon_{\mu\nu\rho\sigma}P^\rho n^\sigma(P).
\end{eqnarray}
Besides, within such power counting, one would have $\mathcal{P}\sim\mathcal{O}(\hbar^2)$, and $\Sigma_P\sim\mathcal{O}(\hbar^2)$, they are thus excluded from the current problem. For later convenience, the zeroth order and first order Wigner functions of massive fermion are given by 
\begin{eqnarray}
S^{<(0)}&=&2\pi\epsilon(P\cdot u)\delta(P^2-m^2)(m+\gamma^\mu P_\mu)f_V(P),\nonumber\\
S^{<(1)}&=&2\pi\epsilon(P\cdot u)\delta(P^2-m^2)\Big(\gamma^5\gamma^\mu n_\mu(P)+\frac{\sigma^{\mu\nu}}{2}\Sigma_{\mu\nu}(P)\Big).
\end{eqnarray}
For massless fermion $S^{<(0)}=2\pi\epsilon(P\cdot u)\delta(P^2)\gamma^\mu P_\mu f_V(P)$ and $S^{<(1)}=2\pi\epsilon(P\cdot u)\delta(P^2)\gamma^5\gamma^\mu n_\mu(P)$. In the following, we use the variable $n_\mu$ instead of $\mathcal{A}_\mu$ for the axial-vector component to avoid the coefficient $2\pi\epsilon(P\cdot u)\delta(P^2-m^2)$ on both sides of the transport equation. 
\section{Coulomb scattering}
\label{s3}
In this section, we consider the scenario where the massive hard fermion probes into a  hot QED plasma and undergoes a 2-by-2 scattering with hot medium at local equilibrium. The light fermion in the medium can be well approximated as massless at local equilibrium. The mass of the probe fermion is assumed to be much greater than the thermal mass $m\gg eT$, in this case the Compton scattering does not contribute at the leading logarithmic order, thus only the Coulomb scattering is considered. This approximation can be understood as a toy model for the spin evolution of the s-quark in the quark gluon plasma. Since the axial-vector component is counted to be $\mathcal{A}_\mu\sim\mathcal{O}(\hbar)$, both the diffusion of spin of the massive probe fermion and the first order gradients of the medium contribute at the same order and should be treated on the same basis. We calculate the collision term of axial kinetic equation keeping all the contributions of the first order gradient and work out the leading logarithmic order collision terms. The following Feynman diagram \cite{Lin:2021mvw} describes the Coulomb scattering of the massive probe fermion ($P$ and $K$) with the massless medium fermion ($P'$ and $K'$).
\begin{figure}[H]\centering
\includegraphics[width=0.3\textwidth]{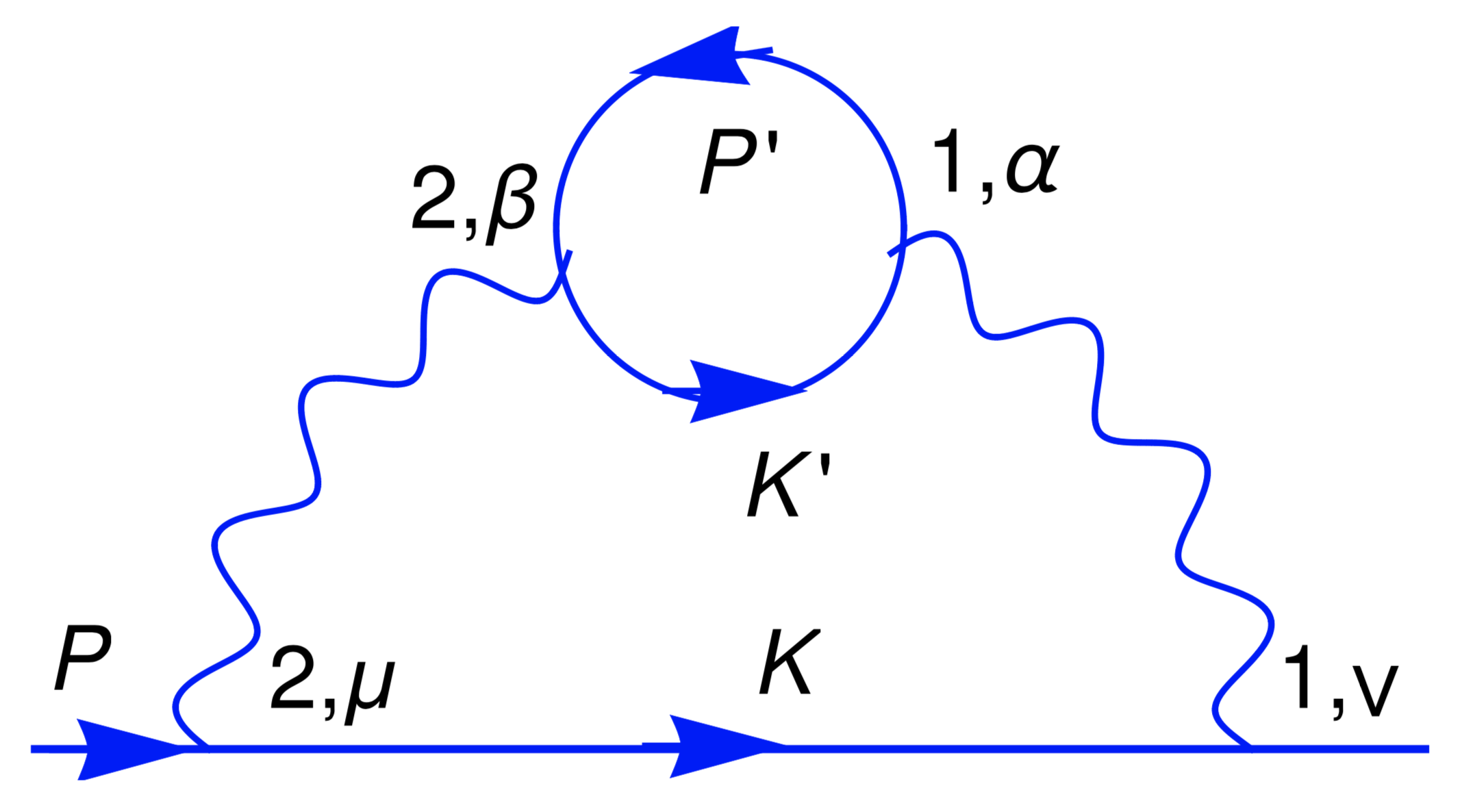}
\caption{Two-loop diagram for fermion self-energy containing propagator corrections \cite{Lin:2021mvw}. the 12 libeling are uniquely determined by the requirement that three propagators attached to a vertex cannot be simultaneously on-shell.}
\label{fig1}
\end{figure}
The greater fermion self-energy is defined as 
\begin{eqnarray}
\Sigma^>(P)=e^4\int_{Q,K}\gamma^\mu S^>(K)\gamma^\nu G_{\mu\nu}^{<}(Q),
\end{eqnarray}
where $\int_{Q,K}=\int d^4Kd^4Q(2\pi)^{-8}\epsilon(K\cdot u)\delta(K^2-m^2)(2\pi)^4\delta(P-K-Q)$. $G_{\mu\nu}^{<}(Q)$ is the photon propagator containing one fermion loop correction, 
\begin{eqnarray}
G^{(0,1)<}_{\mu\nu}(Q)&=&-D_{\mu\beta}^{22}(Q)D_{\alpha\nu}^{11}(Q)\Pi^{(0,1)<\alpha\beta}(Q).
\end{eqnarray}
For simplicity, we will choose Feynman gauge for $D_{\mu\beta}^{22}$ and $D_{\alpha\nu}^{11}$, namely $D_{\mu\beta}^{22}=\frac{ig_{\mu\beta}}{Q^2}$ and $D_{\alpha\nu}^{11}=\frac{-ig_{\alpha\nu}}{Q^2}$. As processes with the on-shell photon such as Compton scattering is suppressed, the collision term is gauge invariant. A short proof of gauge invariance is placed in the Appendix.\ref{gaugeissue}. The photon self-energy get gradient correction from the on-shell loop fermion, it is also counted in $\hbar$ with the leading order and first order photon self-energy are
\begin{eqnarray}
\Pi^{(0)<\alpha\beta}(Q)&=&-e^2\int_{K',P'}\text{Tr}[\gamma^\alpha S^{(0)<}(P')\gamma^\beta S^{(0)>}(K')],\nonumber\\
\Pi^{(1)<\alpha\beta}(Q)&=&-e^2\int_{K',P'}\text{Tr}[\gamma^\alpha S^{(1)<}(P')\gamma^\beta S^{(0)>}(K')+\gamma^\alpha S^{(0)<}(P')\gamma^\beta S^{(1)>}(K')],
\end{eqnarray}
where $\int_{K',P'}=\int d^4K'd^4P'(2\pi)^{-8}\epsilon(P'\cdot u)\epsilon(K'\cdot u)\delta(P'^2)\delta(K'^2)(2\pi)^4\delta(Q+P'-K')$. Fermion with momentum $P$ and $K$ are the massive probe fermion, while with momentum $P'$ and $K'$ are the massless medium fermion. Substituting the Wigner function of the loop fermion, the photon propagator at zeroth order and first order are given by
\begin{eqnarray}
\label{photon_propa}
G^{(0)<}_{\mu\nu}(Q)&=&4e^2(2\pi)^2\int_{K',P'}\frac{1}{(Q^2)^2}\Big(P'_{\{\nu} K'_{\mu\}}-g_{\mu\nu}P'\cdot K'\Big) {f}_V(P')\bar{f}_V(K'),\nonumber\\
G^{(1)<}_{\mu\nu}(Q)&=&
4e^2(2\pi)^2\int_{K',P'}\frac{1}{(Q^2)^2}i\epsilon_{\mu\nu\rho\sigma}\Big(K'^\sigma n^\rho(P')\bar{f}_V(K')-P'^\sigma\bar{n}^\rho(K'){f}_V(P')\Big).
\end{eqnarray}
$G^{(0)<}_{\mu\nu}$ is symmetric in indices, while $G^{(1)<}_{\mu\nu}$ is anti-symmetric. Instead of using the HTL photon propagator and calculate the one loop fermion self-energy \cite{Li:2019qkf, Yang:2020hri}, here we use the loop-corrected photon propagator. The former captures the classical effects in the evolution of probe fermion, while quantum effects such as contributions from the gradients of the medium are not included. One loop fermion self-energy using the HTL photon propagator also assumes that the medium fermions are at equilibrium, non-equilibrium effects are also excluded. In comparison, the zeroth order $G^{(0)}_{\mu\nu}(Q)$ includes classical effects same as described by HTL photon propagator, and through the first order propagator $G^{(1)}_{\mu\nu}(Q)$, spin of the massless medium fermion could contribute to the spin evolution of the probe fermion. Besides, by calculating the two loop fermion self-energy, one can also investigate the evolution of probe fermion in a non-equilibrium medium. However, in this paper, we restrict to the scenario where the massless medium fermion is at local equilibrium as a first step.

Contract $\chi^{<\mu\nu}=\gamma^\mu S^<\gamma^\nu$ with the photon propagator, the fermion self-energy can be decomposed to various Dirac components \cite{Yang:2020hri}, 
\begin{eqnarray}
\label{decomposition}
\chi^{<\mu\nu}G^>_{\mu\nu}
&=&({\mS}G^>_{\mu\nu}g^{\mu\nu}+i{\mS}^{\mu\nu}G^>_{\mu\nu})+i\gamma^5(-{\mP}G^{>\mu}_{\mu}-i{\mS}_{\alpha\beta}G^>_{\mu\nu}\frac{\epsilon^{\mu\nu\alpha\beta}}{2})+\gamma^\rho({\mV}^\mu G^>_{(\mu\rho)}-{\mV}_\rho G^{>\mu}_{\mu}-i\epsilon_{\mu\nu\sigma\rho}{\mA}^\sigma G^{>\mu\nu})\\
&&+\gamma^5\gamma^\rho(-{\mA}^\mu G^>_{(\mu\rho)}+{\mA}_\rho G^{>\mu}_{\mu}+i\epsilon_{\mu\nu\sigma\rho}{\mV}^\sigma G^{>\mu\nu})+\frac{1}{2}\sigma^{\rho\sigma}(2{\mS}^\mu_{~\rho}G^>_{(\mu\sigma)}+{\mS}_{\rho\sigma}G^{>\mu}_{\mu}-2i{\mS}G^>_{\rho\sigma}-i{\mP}\epsilon_{\mu\nu\rho\sigma}G^{>\mu\nu}).\nonumber
\end{eqnarray}
For the purpose of obtaining the collision terms to $\mathcal{O}(\hbar)$, only terms linear in axial-vector component $\mathcal{A}^\mu$ appear while higher order of $\mathcal{A}$ are at least $\mathcal{O}(\hbar^2)$ and are neglected. The self-energy components appearing in the transport equation (\ref{Atransport}) are $\Sigma_S$, $\Sigma_{V\mu}$, $\Sigma_{A\mu}$ and $\Sigma_{T\mu\nu}$, the greater components are expressed as
\begin{eqnarray}
\label{selfenergy_components}
\bar{\Sigma}_S(P)&=&e^2\int_{Q,K}\bar{\mS}(K)G^{(0)<}_{\mu\nu}(Q)g^{\mu\nu},\nonumber\\
\bar{\Sigma}_{V\mu}(P)&=&e^2\int_{Q,K}2\bar{\mV}^{\nu}(K) G^{(0)<}_{\mu\nu}(Q)-\bar{\mV}_\mu(K)G^{(0)<}_{\rho\nu}(Q)g^{\rho\nu},\nonumber\\
\bar{\Sigma}_{A\mu}(P)&=&e^2\int_{Q,K}-2\bar{\mA}^\nu(K) G^{(0)<}_{\mu\nu}(Q)+\bar{\mA}_\mu(K) G^{(0)<}_{\rho\nu}(Q)g^{\rho\nu}+i\epsilon_{\alpha\beta\nu\mu}\bar{\mV}^\nu(K) G^{(1)<\alpha\beta}(Q),\nonumber\\
\bar{\Sigma}_{T\mu\nu}(P)&=&e^2\int_{Q,K}2\bar{\mS}^\rho_{~[\mu}(K)G^{(0)<}_{\rho\nu]}(Q)+\bar{\mS}_{\mu\nu}(K)G^{(0)<}_{\rho\lambda}(Q)g^{\rho\lambda}-2i\bar{\mS}(K)G^{(1)<}_{\mu\nu}(Q).
\end{eqnarray}
The lessor components of the self-energy can be obtained through replacing $S^>$ by $S^<$, and $G^<$ by $G^>$. 

As a first step, in this paper, we consider the case where the massive probe fermion has local equilibrium number distribution, so that we can use the local equilibrium of number distribution $\bar{f}_{K'}\bar{f}_{K}f_{P}{f}_{P'}-f_{K'}f_{K}\bar{f}_{P}\bar{f}_{P'}=0$ in the derivation. For abbreviation, we denote $f_V(P)$ as a general non-equilibrium number distribution, and denote $f_P=f_V^{\text{leq}}(P)=n_F(P)$ as the local equilibrium number distribution. This simplifies the last term $\partial^\sigma\Sigma_{V}^{\nu}(P)$ in (\ref{Atransport}), with $\Sigma_{V}^{\nu}(P)$ defined in (\ref{selfenergy_components}). Using local equilibrium of number distribution, one can convert $\partial^\mu(\bar{f}_{K'}\bar{f}_{K}{f}_{P'})$ into $\partial^\mu f_P$. After straightforward but tedious algebra, the collision terms of axial kinetic equation can be casted into $P\cdot\partial n_\mu=C_{\mu}$,  
\begin{eqnarray}
\label{collision1}
C_{\mu}&=&-4e^4(2\pi)^3\int_{Q,K}\int_{K',P'}\Big\{M^{A1}_{\mu\nu}(\bar{f}_{K}{f}_{P'}\bar{f}_{K'}+{f}_{K}\bar{f}_{P'}{f}_{K'})n^\nu(P)+M^{A2}_{\mu\nu}(f_{P}{f}_{P'}\bar{f}_{K'}+\bar{f}_{P}\bar{f}_{P'}{f}_{K'}){n}^\nu(K)\nonumber\\
&&\qquad\qquad\qquad\qquad\quad\;\;\;+M^{A3}_{\mu\nu}(\bar{f}_{K}\bar{f}_{K'}{f}_{P'}+{f}_{K}{f}_{K'}\bar{f}_{P'})\partial^\nu f_P\;+M^{A4}_{\mu\nu}(f_{P}{f}_{P'}\bar{f}_{K'}+\bar{f}_{P}\bar{f}_{P'}{f}_{K'})\partial^\nu f_K\nonumber\\
&&\qquad\qquad\qquad\qquad\quad\;\;\;+M^{A5}_{\mu\nu}(\bar{f}_{K'}f_{P}\bar{f}_{K}+{f}_{K'}\bar{f}_{P}{f}_{K})n^\nu_{\text{leq}}(P')+M^{A6}_{\mu\nu}({f}_{P'}f_{P}\bar{f}_{K}+\bar{f}_{P'}\bar{f}_{P}{f}_{K}) {n}^\nu_{\text{leq}}(K')\Big\},
\end{eqnarray}
where $\int_{Q,K}\int_{K',P'}=\int(2\pi)^{-16} d^4Qd^4Kd^4K'd^4P'(2\pi)^8\delta(P-K-Q)\delta(Q+P'-K')\epsilon(K\cdot u)\epsilon(P'\cdot u)\epsilon(K'\cdot u)\delta(K^2-m^2)\delta(P'^2)\delta(K'^2)$. The various effective scattering amplitudes in (\ref{collision1}) are given by 
\begin{eqnarray}
\label{amplitudes}
M^{A1}_{\mu\nu}&=&\frac{1}{(Q^2)^2}g_{\mu\nu}\big(-m^2P'\cdot K'+2P\cdot P' K\cdot K'\big)+\{P'\leftrightarrow K'\},\nonumber\\
M^{A2}_{\mu\nu}&=&\frac{1}{(Q^2)^2}\Big(g_{\mu\nu}\big(K\cdot P P'\cdot K' -2K\cdot K' P\cdot P'\big)-K'\cdot P' K_\mu P_\nu-2P\cdot P' Q_\mu K'_\nu+2P\cdot QP'_{\mu} K'_{\nu}+2K\cdot K' P'_{\mu} P_\nu\Big)\nonumber\\
&&+\{P'\leftrightarrow K'\},\nonumber\\
M^{A3}_{\mu\nu}&=&
\frac{1}{(Q^2)^2}\epsilon_{\mu\nu\alpha\beta}K\cdot K'P^{\beta}P'^{\alpha}+\{P'\leftrightarrow K'\},\nonumber\\
M^{A4}_{\mu\nu}&=&-\frac{1}{(Q^2)^2}\Big(
\epsilon_{\mu\nu\alpha\beta}\big(\frac{1}{2}P'\cdot K'P^{\beta}K^{\alpha} 
+P\cdot P'K^{\beta}K'^{\alpha}\big)
+\epsilon_{\lambda\nu\alpha\beta}K'_\mu P^{\alpha}K^{\beta}P'^{\lambda}
\Big)+\{P'\leftrightarrow K'\},\nonumber\\
M^{A5}_{\mu\nu}&=&\frac{2}{(Q^2)^2}\big(m^2Q\cdot K' g_{\mu\nu}-m^2K'_\mu Q_\nu +K\cdot  K' P_\mu  P_\nu -P\cdot  K' P_\mu K_\nu\big),\nonumber\\
M^{A6}_{\mu\nu}&=&\frac{2}{(Q^2)^2}\big(m^2Q\cdot P' g_{\mu\nu}
-m^2P'_\mu Q_\nu
+ K\cdot P' P_\mu P_\nu
-P\cdot  P' P_\mu K_\nu\big), 
\end{eqnarray}
where $\{P'\leftrightarrow K'\}$ denotes exchanging the two momentum in the terms before, namely $f(P',K')+\{P'\leftrightarrow K'\}$ means $g(P',K')+ g(K',P')$, $g$ is an albitrary function.

The first line in collision term (\ref{collision1}) corresponds to the spin diffusion of the probe fermion, which are similar to classical spin relaxation processes \cite{Li:2019qkf, Yang:2020hri}. The last two lines are polarization of probe fermion contributing from the collection motion of the medium as well as spin of the medium fermion. The third line in (\ref{collision1}) describes the polarization effect due to spacetime gradient of $f_V$ of the probe fermion. The last line is the contribution from spin of the medium fermion. As $\mathcal{A}_\mu$ is $\mathcal{O}(\hbar)$ in the power counting, in order to investigate its evolution, it is of key necessity to evaluate all the first order gradient in the collision term. Before going on to present the collision term after momentum integral, we first discuss each part of the collision terms. 
\paragraph{Diffusion}
Terms with $M^{A1}_{\mu\nu}$ and $M^{A2}_{\mu\nu}$ are the spin diffusion terms, these two terms have the same physical meaning as discussed in \cite{Li:2019qkf, Yang:2020hri}, describing the relaxation of probe spin by the QED dynamics. After simplifying the integral measure, the diffusion term can be recasted into 
\begin{eqnarray}
\label{diffusionterm}
C_{A\mu}^{\text{diff}}&=&-4e^4\int_{q_0,q,k'}\big\{M^{A1}_{\mu\nu}(\bar{f}_K{f}_{P'}\bar{f}_{K'}+{f}_{K}\bar{f}_{P'}{f}_{K'})n^\nu(P)+M^{A2}_{\mu\nu}(f_{P}{f}_{P'}\bar{f}_{K'}+\bar{f}_{P}\bar{f}_{P'}{f}_{K'}){n}^\nu(K)\big\}, 
\end{eqnarray}
where $\int_{q_0,q,k'}$ is the abbreviation for $\frac{1}{(2\pi)^5}\int dq_0 d^3qd^3k'\frac{1}{2p'_02k'_02k_0}\delta(p_0-k_0-q_0)\delta(p'_0-k'_0+q_0)$. 
Instead of photon propagator in the HTL approximation, we use the photon propagator with one-loop correction (\ref{photon_propa}). This leads to the same structure in the collision term, while the coefficients will be different by a constant factor compared with \cite{Yang:2020hri}, coming $SU(N)$ of the symmetry of color field in \cite{Yang:2020hri}. In the HTL approximation, the probe fermion and medium fermion are hard fermions $p,k,p',k'\sim T$, while the momentum transfer is soft $eT\ll q_0,q\ll T$. In order to complete the momentum integral and keep the result to leading logarithmic order, the axial-vector components of the out-going probe fermion is expanded in terms of soft momentum $Q$ as
\begin{eqnarray}
n_\mu(K)=n_\mu(P-Q)\simeq n_\mu(P)-Q^\nu\partial_{P_\perp^\nu} n_\mu(P)+\frac{1}{2}Q^\rho\partial_{P_\perp^\rho}Q^\nu\partial_{P_\perp^\nu}n_\mu(P)+\mathcal{O}(Q^3).
\end{eqnarray}
With restriction of the on-shell condition $\delta(P^2-m^2)$, $n_\mu(P)$ is only function of the three-momentum $P_\perp^\mu$, hence the derivative $\partial_{p_0}n_\mu(P)$ can be dropped. And likewise in the expansion of $f_{K}$ and $f_{P'}$ by soft momentum $Q$. The basic strategy is to expand the integrand to $\mathcal{O}(Q^{-2})$, together with the integral measure which is $\mathcal{O}(Q^{2})$, then the momentum integral gives leading logarithmic result. Further details of momentum integral is presented in Appendix.\ref{diffusion_calculation}.
\paragraph{polarization}
The second and third lines in (\ref{collision1}) contain first order gradients through derivatives of number distribution of the probe $\partial^\nu f_P,\;\partial^\nu f_K$ as well as the axial-vector component of the medium fermion $n^\nu_{\text{leq}}(P')$ and $n^\nu_{\text{leq}}(K')$. We take the assumption that the number distribution of the probe has reached local equilibrium with the medium, in this scenario, we can decompose $\partial^\nu f_P,\;\partial^\nu f_K$ in terms of gradients of the fluid. Similar to \cite{Hidaka:2017auj}, derivative of the fluid velocity $u$ can be decomposed into anti-symmetric and symmetric part $\partial^\mu u^\nu=\omega^{\mu\nu}+\sigma^{\mu\nu}$, where $\omega^{\mu\nu}=(\partial^\mu u^\nu-\partial^\nu u^\mu)/2$ and $\sigma^{\mu\nu}=(\partial^\mu u^\nu+\partial^\nu u^\mu)/2$. With the vorticity defined as $\omega^\mu\equiv\frac{1}{2}\epsilon^{\mu\nu\alpha\beta}u_\nu(\partial_\alpha u_\beta)$, the symmetric part is casted into $\omega_{\alpha\beta}=-\epsilon_{\alpha\beta\mu\nu}\omega^\mu u^\nu+\kappa_{\alpha\beta}$, with $\kappa_{\alpha\beta}=\frac{1}{2}(u_\alpha D u_\beta-u_\beta D u_\alpha)$. Defining $\tilde{E}^\sigma(P)=P_\lambda (\frac{1}{T}u^\lambda\partial^\sigma T-\sigma^{\sigma\lambda}-\kappa^{\sigma\lambda})$, we arrive at the following decomposition,
\begin{eqnarray}
\label{decomposition}
\partial^\nu f_P=-\big(
\epsilon^{\nu\rho\alpha\beta} P_\rho \omega_\alpha u_\beta+\tilde{E}^\nu(P)\big)f'_P,
\end{eqnarray}
where $f'_P=\partial_{u\cdot P} f_P =(-\frac{1}{T})f_P\bar{f}_P$. $\tilde{E}^\sigma(P)$ can be further casted in combination of shear tensor, acceleration and gradient of temperature 
\begin{eqnarray}
\tilde{E}^\sigma(P)= -P_\lambda(\partial^{\langle\sigma}u^{\lambda\rangle}+\frac{1}{3}\Delta^{\sigma\lambda}\theta+u^{\sigma}D u^{\lambda})+P\cdot u[\partial^\sigma \ln T]. 
\end{eqnarray}
The spin evolution of the massive probe fermion in a massless QED at local equilibrium also involves the exchanging of spin with the massless medium fermion, which are at local equilibrium. For the medium fermion, we take the local equilibrium distribution of spin \cite{{Hidaka:2017auj}}, namely 
\begin{eqnarray}
\label{masslesseq}
\mathcal{A}_\mu^{\text{leq}}(P)
&=&2\pi\epsilon(P\cdot u)\delta(P^2)\Big(
\frac{P\cdot\omega\,u_\mu }{2}  
-\frac{P\cdot u\omega_\mu }{2}
-S^{(u)}_{\mu\sigma}(P)\tilde{E}^\sigma(P)\Big)f'_V(P),
\end{eqnarray}
where 
\begin{eqnarray}
S^{(u)}_{\mu\nu}(P)=\frac{\epsilon_{\mu\nu\alpha\beta}P^\alpha u^\beta}{2P\cdot u}.
\end{eqnarray}
The polarization effect contains contribution from vorticity, shear tensor, acceleration and the gradient of temperature. With the decomposition (\ref{decomposition}) and (\ref{masslesseq}), the vorticity related terms in the collision term can be collected into, 
\begin{eqnarray}
\label{collision-vorticity}
C_{A\mu}^{\text{vor}}&=&-4e^4\int_{q_0,q,k'}C_{\mu}^{\text{vor}}(-\beta)\bar{f}_{K}\bar{f}_{K'}{f}_{P'}f_{P},
\end{eqnarray}
with $\beta(x)=T(x)^{-1}$, and 
\begin{eqnarray}
\label{tensors-vorticity}
C_{\mu}^{\text{vor}}&=&-(M^{A3}_{\mu\nu}P_\rho+M^{A4}_{\mu\nu}K_\rho ) \epsilon^{\nu\rho\alpha\beta} \omega_\alpha u_\beta+\frac{1}{2}(M^{A5}_{\mu\nu}P'_\rho+M^{A6}_{\mu\nu}K'_\rho)\omega^{[\rho} u^{\nu]}.
\end{eqnarray}
The effective amplitudes $M^{Ai}$ defined in (\ref{amplitudes}). (\ref{collision-vorticity}) characterizes the polarization effect due the the vorticity in the medium. Using the decomposition (\ref{decomposition}) and (\ref{masslesseq}), the shear tensor related terms in the collision term can be casted into, 
\begin{eqnarray}
\label{collision-shear}
C_{A\mu}^{\text{shear}}&=&-4e^4\int_{q_0,q,k'}C_{\mu\alpha\beta}^{\text{shear}}\sigma^{\langle\alpha\beta\rangle}(-\beta)\bar{f}_{K}\bar{f}_{K'}{f}_{P'}f_{P}
\end{eqnarray}
with $C_{\mu\alpha\beta}^{\text{shear}}$ defined through
\begin{eqnarray}
\label{C-shear}
C_{\mu\alpha\beta}^{\text{shear}}=M^{A3}_{\mu\alpha}P_\beta  +M^{A4}_{\mu\alpha}K_\beta+\frac{1}{2P'\cdot u}(M^{A5})_{\mu}^{~\nu}\epsilon_{\nu\alpha\rho\lambda}P'^\rho u^\lambda
P'_\beta +\frac{1}{2K'\cdot u}(M^{A6})_{\mu}^{~\nu} \epsilon_{\nu\alpha\rho\lambda}K'^\rho u^\lambda K'_\beta. 
\end{eqnarray}
The shear tensor $\sigma^{\langle\alpha\beta\rangle}$ is the symmetric and traceless part of $\sigma^{\alpha\beta}=\frac{1}{2}(\partial_\perp^\alpha u^\beta+\partial_\perp^\beta u^\alpha)-\frac{1}{3}\Delta^{\alpha\beta}\theta$. In the local rest frame of the fluid, the shear tensor only spatial components $\sigma_{ij}=\frac{1}{2}(\partial_i u_j+\partial_j u_i)-\frac{1}{3}\delta_{ij}\vec\partial\cdot \vec u$. The efficient amplitudes are presented in (\ref{amplitudes}). The remaining first order gradients are the temperature gradient and acceleration, the corresponding collision term is collected into
\begin{eqnarray}
\label{collision_Tgra_acc}
C_{A\mu}^{\text{Tgra+acc}}&=&-4e^4\int_{q_0,q,k'}\big(C_{\mu\lambda}^{\text{Tgra}}\partial^\lambda\ln T+ C_{\mu\lambda}^{\text{acc}}D u^\lambda\big)
(-\beta)\bar{f}_{K}\bar{f}_{K'}{f}_{P'}f_{P},
\end{eqnarray}
with coefficients $C_{\mu\lambda}^{\text{Tgra}}$ and $C_{\mu\lambda}^{\text{acc}}$ defined as
\begin{eqnarray}
\label{C_Tgra_acc}
C_{\mu\lambda}^{\text{Tgra}}&=&-M^{A3}_{\mu\lambda}P\cdot u
-M^{A4}_{\mu\lambda}K\cdot u
-\frac{1}{2}(M^{A5})_\mu^{\;\,\nu}\epsilon_{\nu\lambda\alpha\beta}P'^\alpha u^\beta
-\frac{1}{2}(M^{A6})_\mu^{\;\,\nu}\epsilon_{\nu\lambda\alpha\beta}K'^\alpha u^\beta,\nonumber\\
C_{\mu\lambda}^{\text{acc}}&=&M^{A3}_{\mu\nu}u^\nu P_\lambda +M^{A4}_{\mu\nu}u^\nu K_\lambda.
\end{eqnarray}
The final result of the collision term will be the sum of the all the parts above, namely $C_{A\mu}=C_{A\mu}^{\text{diff}}+C_{A\mu}^{\text{vor}}+C_{A\mu}^{\text{shear}}+C_{A\mu}^{\text{Tgra+acc}}$. The further calculation of the collision terms is presented in Appendix.\ref{firstorder_calculation}. 
\section{result}
\label{s4}
In this section, we show explicitly the result of the collision term. The leading logarithmic contribution comes from the soft $eT\ll q_0, q\ll T$ regime, the basic strategy to obtain the leading logarithmic contribution is to collecte all the terms up to $\mathcal{O}(Q^{-2})$ in the integrand. Combined with the measure which is $\mathcal{O}(Q^{2})$, both combined will give the leading logarithmic results. With the assumption that mass of the probe fermion is much larger than thermal mass $m\gg m_D\sim eT$, Compton scattering is sub-leading and only Coulomb scattering is considered. In the calculation, there is no more restriction for the mass of the probe fermion. In the following, we present the collision term after the momentum integral for arbitrary mass of the probe fermion, and also take massless and non-relativistic limit for a comparison. 
\subsection{arbitrary mass}
For arbitrary nonzero mass of the probe fermion, the kinetic equation of the axial-vector component becomes
\begin{eqnarray}
\label{general}
P\cdot \partial n^\mu(P)&=&-\kappa_{LL}\frac{T}{m v}\Big\{C^{(1)} n^\mu(P)
+C^{(2)} u^\mu
+C^{(3)} \hat{P}_\perp^\mu
+C^{(4)} \hat{P}_\perp^\nu \partial_{P_{\perp\mu}}n_\nu(P)\nonumber\\
&&\qquad\qquad
+C^{(5)} \hat{P}_\perp^\nu \partial_{P_\perp^\nu}n^\mu(P)
+C^{(6)} g^{\nu\rho} \partial_{P_\perp^\nu}\partial_{P_\perp^\rho}n^\mu(P)
+C^{(7)} \hat{P}_\perp^\nu\hat{P}_\perp^\rho\partial_{P_\perp^\nu}\partial_{P_\perp^\rho}n^\mu(P)\nonumber\\
&&\qquad\qquad
+C^{(8)}(\omega^\mu+\hat{P}_\perp^\mu\hat{P}_\perp^\nu\omega_\nu)
+C^{(9)}\frac{1}{2}\Big(\epsilon^{\mu\nu\rho\alpha}u_\nu \hat{P}_{\perp\rho} \hat{P}_{\perp}^\beta+\epsilon^{\mu\nu\rho\beta}u_\nu  \hat{P}_{\perp\rho}  \hat{P}_{\perp}^\alpha\Big)\sigma_{\langle\alpha\beta\rangle}\nonumber\\
&&\qquad\qquad
+C^{(10)}\epsilon^{\mu\nu\alpha\beta}u_\alpha\hat{P}_{\perp\beta}Du_\nu
+C^{(11)}\epsilon^{\mu\nu\alpha\beta}u_\alpha\hat{P}_{\perp\beta}\partial_\nu\ln T
\Big\},
\end{eqnarray}
note that both sides of the kinetic equation are on the mass shell $\delta(P^2-m^2)$. $\kappa_{LL}$ is the leading logarithmic coefficient defined by 
\begin{eqnarray}
\kappa_{LL}=e^4\ln\frac{1}{e}\frac{T^2}{8\pi}.
\end{eqnarray}
Similar to \cite{Yang:2020hri}, we also introduce the four velocity $v^\mu\equiv P^\mu/m$ for simplicity. Which has the normalization $v^\mu v_\mu=1$, and $v_0=p_0/m$ with $v=|\vec{p}|/m$. The rapidity is $\eta_p\equiv\text{arctanh}(p/p_0)\equiv2^{-1}\ln[(p_0+p)/(p_0-p)]$, also define for simplicity $\theta_n\equiv v-v_0^n\eta_p$. The coefficients in (\ref{general}) are
\begin{eqnarray}
\label{coe_general}
C^{(1)} &=&\frac{v}{3 v_0}-\frac{m^2 v_0\theta _{-1}}{3 T^2} (1-f_p) f_p-\frac{m v^3 }{3 T v_0^2}(1-2 f_p)\nonumber\\
C^{(2)} &=&\Big(\frac{1}{3}+\frac{\theta_{-1}}{3 v}-\frac{m v_0 \theta_{-1} }{6 T v}(1-2 f_p)\Big)\hat{P}_\perp^\nu  n_\nu(P)+\frac{m (2 v^3-v_0^2\theta_{-1} )}{6 v^2}\partial_{P_{\perp}^\nu}n^\nu(P)
+\frac{m(2v^3-3v_0^2\theta_{-1})}{6 v^2}\hat{P}_\perp^\rho\hat{P}_\perp^\nu \partial_{P_\perp^\rho} n_\nu(P),\nonumber\\
C^{(3)} &=&\Big(\frac{v}{3 v_0^3}+\frac{\theta_{-1}}{3 v_0}-\frac{m\theta_{-1}}{6 T}(1-2 f_p)\Big)\hat{P}_\perp^\nu  n_\nu(P)-\frac{ m v_0\theta_{-1}}{6 v}\partial_{P_{\perp}^\nu}n^\nu(P)
+\frac{m (2 v^3-3 v_0^2\theta_{-1} )}{6 v v_0}\hat{P}_\perp^\rho\hat{P}_\perp^\nu \partial_{P_\perp^\rho} n_\nu(P),\nonumber\\
C^{(4)} &=&\frac{m v^2}{3 v_0},\nonumber\\
C^{(5)} &=&\frac{m}{3 v_0}-\frac{m^2 v_0^2\theta_{-1}  }{6 T v}(1-2 f_p),\nonumber\\
C^{(6)} &=&\frac{m^2 (3 v^3 v_0- v_0^3\theta_{-3})}{12 v^2},\nonumber\\
C^{(7)} &=&-\frac{m^2 v_0(2 \theta_1+ \theta_{-1} )}{12 v^2},\nonumber\\
C^{(8)} &=&\Big(-\frac{m v}{3 v_0^2}-\frac{m^2 v_0 \theta _{-1} }{6 T}(1-2 f_p)\Big)\frac{(1-f_p) f_p }{2 T},\nonumber\\
C^{(9)} &=&\Big(\frac{m (v^5+3 v_0^2 \theta _1)}{3 v^2 v_0^2}+\frac{m^2 v_0 (2 v^3-(v^2+3 v_0^2)\theta _{-1} )}{6 T v^2}(1-2 f_p)\Big)\frac{(1-f_p) f_p }{2 T},\nonumber\\
C^{(10)} &=&\Big(\frac{m v_0\theta_{-1} }{2 v}+\frac{m^2  (2 v^3-3 \theta _{-1} v_0^2)}{12 T v}(1-2 f_p)\Big)\frac{(1-f_p) f_p }{2 T},\nonumber\\
C^{(11)} &=&\Big(\frac{m (2 v-3 v_0^2\theta_{-1})}{6 v v_0}+\frac{m^2 (3 v^3+5 \theta _1)}{12 T v}(1-2 f_p)\Big)\frac{(1-f_p) f_p }{2 T}.
\end{eqnarray}
$f_p$ is the local equilibrium number distribution of the probe, which is the Fermi Dirac distribution $f_p=1/(e^{E_p/T}+1)$, and $E_p=\sqrt{p^2+m^2}$.

The first two lines in (\ref{general}) are the spin-diffusion terms. These terms has same structure as \cite{Yang:2020hri}, while the explicit coefficients different by a $SU(N)$ group constant. The last two lines in (\ref{general}) are the polarization effect induced by the first order gradients of the medium, which include vorticity, shear tensor, acceleration and gradient of temperature. If the probe fermion is not polarized in the initial state, there will be polarization effects during the evolution, until the polarization effect balance the diffusion effect, then the spin of the probe fermion will achieve local equilibrium of the medium. It is worth emphasizing that, we have taken the assumption that the number distribution of the probe fermion has achieved local equilibrium with the medium. In general, the relaxation of $f_V$ of the probe is coupled with the relaxation of spin $n^\mu$ and should also be incorporated. These will be presented in further studies.  
\subsection{massless limit}
Generally, when $m\ll eT$ is not satisfied, the Compton scattering with polarized photon is no more suppressed and contributes also at leading logarithmic order. However, the Compton scattering is not included the current paper, the massless limit is considered to compare with other researches. In the massless limit, the structure of the collision term is unchanged, the only difference lies in the replacement in the coefficients, with $C^{(i)}$ replaced by $C^{(i)}_{\text{chi}}$. With $C^{(7)}_{\text{chi}} =0$, the non vanishing coefficients are,
\begin{eqnarray}
\label{coe_massless}
C^{(1)}_{\text{chi}} &=&\frac{1}{3}-\frac{p^2 }{3 T^2} f_p(1-f_p)-\frac{p }{3 T}(1-2 f_p),\nonumber\\
C^{(2)}_{\text{chi}} &=&\big(\frac{2}{3}-\frac{p }{6 T}(1-2 f_p)\big)\hat{P}_\perp^\nu  n_\nu(P)+\frac{p}{6}\partial_{P_{\perp}^\nu}n^\nu(P)
-\frac{p}{6}\hat{P}_\perp^\rho\hat{P}_\perp^\nu \partial_{P_\perp^\rho} n_\nu(P),\nonumber\\
C^{(3)}_{\text{chi}} &=&\big(\frac{1}{3}-\frac{p }{6 T}(1-2 f_p)\big)\hat{P}_\perp^\nu  n_\nu(P)-\frac{p}{6}\partial_{P_{\perp}^\nu}n^\nu(P)
-\frac{p}{6}\hat{P}_\perp^\rho\hat{P}_\perp^\nu \partial_{P_\perp^\rho} n_\nu(P),\nonumber\\
C^{(4)}_{\text{chi}} &=&\frac{p}{3},\qquad \qquad
C^{(5)}_{\text{chi}} ~=~-\frac{p^2 }{6 T}(1-2 f_p),\qquad\qquad
C^{(6)}_{\text{chi}} ~=~\frac{p^2}{6},\nonumber\\
C^{(8)}_{\text{chi}} &=&-\frac{p^2 (1-2 f_p) }{6T}\frac{f_p(1-f_p)}{2T},\qquad \qquad\qquad\qquad\quad\,
C^{(9)}_{\text{chi}} ~=~\big(-\frac{p}{3}+\frac{p^2}{3T}(1-2f_p)\big)\frac{f_p(1-f_p)}{2T},\nonumber\\
C^{(10)}_{\text{chi}}&=&\big(\frac{p}{2}-\frac{p^2 (1-2 f_p)}{12 T}\big)\frac{f_p(1-f_p)}{2T},\qquad \qquad\qquad\;\;
C^{(11)}_{\text{chi}}~=~\big(-\frac{p}{2}+\frac{p^2 (1-2 f_p)}{4 T}\big)\frac{f_p(1-f_p)}{2T}.
\end{eqnarray}

\subsection{non-relativistic limit}
To investigate the spin evolution of heavy quark in the quark gluon plasma, the non-relativistic limit of the kinetic equation is analyzed. In the non-relativistic limit, it is assumed that $m\gg p, T$. We keep the collision term to $\mathcal{O}(m^{-2})$ for later convenience when checking the eliminating of the collision term in the global equilibrium. 
\begin{eqnarray}
\label{nonrelativistic}
\Big(\partial_t+\frac{1}{m}P_\perp^\nu\partial_\nu\Big) n^\mu(P)&=&-\kappa_{LL}\Big\{C^{(1)}_{\text{non}} n^\mu(P)
+C^{(2)}_{\text{non}} u^\mu
+C^{(3)}_{\text{non}} \hat{P}_\perp^\mu
+C^{(4)}_{\text{non}} \hat{P}_\perp^\nu \partial_{P_{\perp\mu}}n_\nu(P)\nonumber\\
&&\qquad\;
+C^{(5)}_{\text{non}} \hat{P}_\perp^\nu \partial_{P_\perp^\nu}n^\mu(P)
+C^{(6)}_{\text{non}} g^{\nu\rho} \partial_{P_\perp^\nu}\partial_{P_\perp^\rho}n^\mu(P)
\nonumber\\
&&\qquad\;
+C^{(8)}_{\text{non}}(\omega^\mu+\hat{P}_\perp^\mu\hat{P}_\perp^\nu\omega_\nu)
+C^{(9)}_{\text{non}}\frac{1}{2}\Big(\epsilon^{\mu\nu\rho\alpha}u_\nu \hat{P}_{\perp\rho} \hat{P}_{\perp}^{\beta}+\epsilon^{\mu\nu\rho\beta}u_\nu  \hat{P}_{\perp\rho}  \hat{P}_{\perp}^{\alpha}\Big)\sigma_{\langle\alpha\beta\rangle}
\nonumber\\
&&\qquad\;
+C^{(10)}_{\text{non}}\epsilon^{\mu\nu\alpha\beta}u_\alpha\hat{P}_{\perp\beta}Du_\nu
+C^{(11)}_{\text{non}}\epsilon^{\mu\nu\alpha\beta}u_\alpha\hat{P}_{\perp\beta}\partial_\nu\ln T
\Big\},
\end{eqnarray}
with the coefficients are 
\begin{eqnarray}
\label{coe_non}
C^{(1)}_{\text{non}} &=&\frac{1}{m^2}\big(\frac{T}{3}-\frac{2 p^2  }{9 T}f_p(1-f_p)\big),\qquad\qquad\quad\;\;
C^{(2)}_{\text{non}} ~=~\big(\frac{T}{3m p}-\frac{p }{9 m^2}(1-2 f_p)\big)\hat{P}_\perp^\nu  n_\nu(P)+\frac{2T}{9m}\partial_{P_{\perp}^\nu}n^\nu(P),\nonumber\\
C^{(3)}_{\text{non}} &=&\frac{T}{3m^2}\hat{P}_\perp^\nu  n_\nu(P)-\frac{T p}{9 m^2}\partial_{P_{\perp}^\nu}n^\nu(P),
\qquad\;\;\,
C^{(4)}_{\text{non}} ~=~\frac{T p}{3m^2},\nonumber\\
C^{(5)}_{\text{non}} &=&\frac{T}{3 p}-\frac{p }{9 m}(1-2 f_p)-\frac{p\,T}{6 m^2},\qquad\qquad\quad
C^{(6)}_{\text{non}} ~=~\frac{T}{9}-\frac{p^2 T}{12 m^2},\nonumber\\
C^{(8)}_{\text{non}} &=&\big(-\frac{T}{3 m}-\frac{p^2}{9 m^2}\big)\frac{ f_p(1-f_p) }{2 T},\qquad\qquad\;\;\,
C^{(9)}_{\text{non}} ~=~\big(-\frac{T}{3 m}-\frac{4 p^2 (1-2 f_p)}{9 m^2}\big)\frac{ f_p(1-f_p) }{2 T},\nonumber\\
C^{(10)}_{\text{non}} &=&\frac{p T}{3 m^2}\frac{ f_p(1-f_p) }{2 T},\qquad\qquad
\qquad\qquad\quad\;\;
C^{(11)}_{\text{non}} ~=~\big(\frac{T}{3 p}+\frac{p (1-2 f_p)}{9 m}-\frac{p T}{2 m^2}\big)\frac{ f_p(1-f_p) }{2 T}.
\end{eqnarray}
Both diffusion and polarization get suppressed, guarantees that in the heavy quark limit $m\rightarrow\infty$, the orientation of spin is fixed while the spin density still experience the diffusion process. The different behavior in spin orientation and spin density can be observed more clearly when further decomposing kinetic equation of $n^\mu$ to its three degrees of freedom, as presented in Sec.\ref{s5}.

\subsection{Relaxation near global equilibrium}
For quantum kinetic theory, the elimination of collision term in global equilibrium for massive fermion has been proved in \cite{Wang:2020pej, Weickgenannt:2020aaf}. The collision terms in chiral kinetic theory is also shown to be vanishing in local equilibrium \cite{Hidaka:2017auj, Fang:2022ttm}. We here check the vanishing of collision term in global equilibrium as a guarantee of the correctness the above calculation, and also as a prerequisite for extracting the relaxation rate. In global equilibrium, $n_\mu(P)$ in a purely rotating fluid could be defined frame-independently \cite{Wang:2020pej,Hidaka:2017auj},
\begin{eqnarray}
\label{n_geq}
n_\mu^\text{geq}(P)&=&\Big(\frac{P\cdot\omega\,u_\mu }{2} -\frac{P\cdot u\omega_\mu}{2}\Big)f'_P. 
\end{eqnarray}
Using the following derivatives of the equilibrium number distribution function, $f_p=n_F(E_p)$,
\begin{eqnarray}
\partial_{P_\perp^\mu}f_p&=&f_p(1-f_p)\frac{P_{\perp\mu}}{p_0T},\nonumber\\
\partial_{P_\perp^\mu}\partial_{P_\perp^\nu}f_p&=&f_p(1-f_p)(1 - 2 f_p)\frac{P_{\perp\mu}P_{\perp\nu}}{p_0^2T^2}+f_p(1-f_p)\frac{1}{p_0T}\Big(\frac{P_{\perp\mu}P_{\perp\nu}}{p_0^2}+\Delta_{\mu\nu}\Big),
\end{eqnarray}
together with the following tensors for the momentum derivatives of $n_\mu^\text{geq}(P)$,
\begin{eqnarray}
\partial_{P_{\perp}^\nu}n_\mu^\text{geq}(P)&=&-\frac{f_p(1-f_p)}{2T}\Big(\omega_\nu u_\mu+\frac{P_{\perp\nu}}{p_0}\omega_\mu\Big)-\frac{f_p(1-f_p)(1-2f_p)}{2T}\frac{P_{\perp\nu}}{p_0 T}\big(P_\perp^\sigma\omega_\sigma u_\mu-p_0\omega_\mu\big),
\nonumber\\
\partial_{P_{\perp}^\nu}\partial_{P_{\perp}^\rho}n_\mu^\text{geq}(P)&=&
-\frac{f_p(1-f_p)}{2T}\frac{1}{p_0}\Big(\Delta_{\nu\rho}+\frac{P_{\perp\nu}P_{\perp\rho}}{p_0^2}\Big)\omega_\mu\nonumber\\
&&-\frac{f_p(1-f_p)(1-2f_p)}{2T}\frac{1}{p_0T}\Big(\big(\Delta_{\nu\rho}+\frac{P_{\perp\nu}P_{\perp\rho}}{p_0^2}\big)\big(P_\perp^\sigma\omega_\sigma u_\mu-p_0\omega_\mu\big)+\big(P_{\perp(\rho}\omega_{\nu)}u_\mu+\frac{2P_{\perp\nu}P_{\perp\rho}}{p_0}\omega_\mu\big)\Big)\nonumber\\
&&-\frac{f_p(1-f_p)\big((1-2f_p)^2-2f_p(1-f_p)\big)}{2T}\frac{P_{\perp\nu}P_{\perp\rho}}{p_0^2T^2}\big(P_\perp^\sigma\omega_\sigma u_\mu-p_0\omega_\mu\big),
\end{eqnarray}
one can easily get all the derivatives in the collision terms. Substituting these derivatives back into the diffusion part, one can explicitly find that the diffusion part balances the vorticity part, and thus the collision terms are eliminated in global equilibrium. The vanishing of collision terms is also be proved for the massless limit and non-relativistic limit when inserting the derivatives in corresponding limits. 

Near the global equilibrium, the relaxation of the spin is dominated by the diffusion terms, leading to the relaxation rate near global equilibrium
\begin{eqnarray}
\label{RTA}
P\cdot \partial n^\mu(P)&=&(\hat{\tau}^{-1})^{\mu\nu}\delta n_\nu(P), 
\end{eqnarray}
where the relaxation time is now an operator, 
\begin{eqnarray}
(\hat{\tau}^{-1})^{\mu\nu}&=&\kappa_{LL}\frac{T}{m v}\bigg\{g^{\mu\nu}\bigg(\frac{v}{3 v_0}-\frac{m^2 v_0\theta _{-1}}{3 T^2} (1-f_p) f_p-\frac{m v^3 }{3 T v_0^2}(1-2 f_p)+\Big(\frac{m}{3 v_0}-\frac{m^2 v_0^2\theta_{-1}  }{6 T v}(1-2 f_p)\Big)\hat{P}_\perp^\alpha \partial_{P_\perp^\alpha}
\nonumber\\
&&\qquad\qquad\qquad+\frac{m^2 (3 v^3 v_0- v_0^3\theta_{-3})}{12 v^2}g^{\alpha\beta} \partial_{P_\perp^\alpha}\partial_{P_\perp^\beta}
-\frac{m^2 v_0(2 \theta_1+ \theta_{-1} )}{12 v^2}\hat{P}_\perp^\alpha\hat{P}_\perp^\beta\partial_{P_\perp^\alpha}\partial_{P_\perp^\beta}\bigg)+\frac{m v^2}{3 v_0}\hat{P}_\perp^\nu \partial_{P_{\perp\mu}} 
\nonumber\\
&&\qquad\quad+\Big(\frac{1}{3}+\frac{\theta_{-1}}{3 v}-\frac{m v_0 \theta_{-1} }{6 T v}(1-2 f_p)\Big)u^\mu\hat{P}_\perp^\nu  +\frac{m (2 v^3-v_0^2\theta_{-1} )}{6 v^2}u^\mu\partial_{P_{\perp}^\nu}
+\frac{m(2v^3-3v_0^2\theta_{-1})}{6 v^2}u^\mu\hat{P}_\perp^\nu \hat{P}_\perp^\rho\partial_{P_\perp^\rho}\nonumber\\
&&\qquad\quad+\Big(\frac{v}{3 v_0^3}+\frac{\theta_{-1}}{3 v_0}-\frac{m\theta_{-1}}{6 T}(1-2 f_p)\Big)\hat{P}_\perp^\mu\hat{P}_\perp^\nu  -\frac{ m v_0\theta_{-1}}{6 v}\hat{P}_\perp^\mu\partial_{P_{\perp}^\nu}
+\frac{m (2 v^3-3 v_0^2\theta_{-1} )}{6 v v_0}\hat{P}_\perp^\mu\hat{P}_\perp^\nu \hat{P}_\perp^\rho\partial_{P_\perp^\rho} \bigg\}.
\end{eqnarray}
The assumption that the number distribution of the probe fermion has reached local equilibrium with the medium leads to the disappearance of gradient terms in the relaxation time. Otherwise, the kinetic equation of both spin and number distribution will couple with each other, the relaxation of the charge also contributes to spin evolution \cite{Wang:2021qnt, Fang:2022ttm}. The full set of kinetic equation of both spin and charge to the first order of gradient will be included in an upcoming paper. 
\section{Numerical Analysis}
\label{s5}
\subsection{Decomposition of kinetic equation}
In the end, we give some preliminary attempts to solve the transport equation. So far, the axial component $\mathcal{A}_\mu$ has been treated as an independent variable in the transport equation, only constrained by the perpendicular relation that $P^\mu \mathcal{A}_\mu=0$. The decomposition of $\mathcal{A}_\mu$ was not considered for the simplicity when deriving the transport equation. Yet, the decomposition is necessary to obtain components with definite physical meanings, and to compare with the chiral limit. Limited by $P^\mu \mathcal{A}_\mu=0$, $\mathcal{A}_\mu$ has three independent degrees of freedom in the massive case, while in the chiral limit, the orientation of spin is locked to the momentum, so that $\mathcal{A}_\mu$ has only one degree of freedom. To get the transport equation for each of the three degrees of freedom of $\mathcal{A}_\mu$ and meanwhile keep the correct massless limit, we adopt the following decomposition of $\mathcal{A}_\mu=2\pi\epsilon(P\cdot u)\delta(P^2-m^2)n_\mu$ \cite{Sheng:2020oqs}, 
\begin{eqnarray}
\label{decompo_n}
n^\mu=P^\mu f_A+\frac{P^2}{(u\cdot P)^2-P^2}{P}_\perp^\mu f_A+\frac{P^2}{u\cdot P}\mathcal{M}_\perp^\mu+\frac{1}{2(u\cdot P)}\epsilon^{\mu\nu\alpha\beta}P_\nu  u_\alpha \partial_\beta f_V.
\end{eqnarray}
$f_A=u\cdot n/u\cdot P$ is identified as the axial-charge density, $\mathcal{M}_\perp^\mu$ is the transverse magnetic dipole-moment $\mathcal{M}_\perp^\mu=\Xi^{\mu\nu}\mathcal{M}_\nu$, where $\mathcal{M}^\mu=-\frac{1}{2}\epsilon^{\mu\nu\alpha\beta}u_\nu \Sigma_{\alpha\beta}$ and $\Sigma_{\alpha\beta}$ is the dipole moment tensor defined in last equation in (\ref{Wignercompo}). With such decomposition, the various components of $n^\mu$ have clear physical meanings and smooth massless limit. See \cite{Sheng:2020oqs, Guo:2020zpa} for more details of the decomposition. 

Consider that the charge distribution has achieved local equilibrium, the general $f_V$ in (\ref{decompo_n}) is replaced by the local equilibrium distribution function. For convenience, we consider the scenario that the medium is in the global equilibrium, namely we keep only the vorticity and neglect all other first order gradients. The transport equations of the axial-charge density $f_A(P)$ and transverse magnetic dipole-moment $\mathcal{M}_\perp^\mu(P)$ become, 
\begin{eqnarray}
\label{fA_M}
P\cdot \partial f_A&=&-\kappa_{LL}\frac{T}{m v}\Big\{C^{(1)}_A f_A
+C^{(2)}_A \hat{P}_\perp^\nu \partial_{P_{\perp}^\nu}f_A
+C^{(3)}_A g^{\nu\rho} \partial_{P_\perp^\nu}\partial_{P_\perp^\rho}f_A
+C^{(4)}_A \hat{P}_\perp^\nu\hat{P}_\perp^\rho\partial_{P_\perp^\nu}\partial_{P_\perp^\rho}f_A\nonumber\\
&&\qquad\qquad+C^{(5)}_A \partial_{P_\perp^\nu}\mathcal{M}_\perp^\nu+C^{(6)}_A \hat{P}_\perp^\nu\omega_\nu\Big\},\nonumber\\
P\cdot \partial \mathcal{M}_\perp^\mu&=&-\kappa_{LL}\frac{T}{m v}\Big\{C^{(1)}_M \mathcal{M}_\perp^\mu
+C^{(2)}_M \hat{P}_\perp^\nu \partial_{P_{\perp}^\nu}\mathcal{M}_\perp^\mu
+C^{(3)}_M \big(g^{\nu\rho} \partial_{P_\perp^\nu}\partial_{P_\perp^\rho}\mathcal{M}_\perp^\mu-\frac{2}{m v}\hat{P}_\perp^\mu \partial_{P_\perp^\nu}\mathcal{M}_\perp^\nu\big)\nonumber\\
&&\qquad\qquad
+C^{(4)}_M \hat{P}_\perp^\nu\hat{P}_\perp^\rho\partial_{P_\perp^\nu}\partial_{P_\perp^\rho}\mathcal{M}_\perp^\mu
+C^{(5)}_M \Xi^{\mu\nu}\partial_{P_{\perp}^\nu}f_A+C^{(6)}_M \Xi^{\mu\nu}\omega_\nu\Big\},
\end{eqnarray}
both equations are coupled, with the coefficient $C_A^{(i)}$ defined as 
\begin{eqnarray}
C_A^{(1)}&=&-\frac{m^2 v_0\theta_{-1} }{3 T^2}(1-f_p) f_p -\frac{ m (2 v^5- v_0^2\theta _{-1})}{6 T v^2 v_0^2}(1-2 f_p)-\frac{v^3+ v_0^2 (v_0^2+1)\theta _1}{3 v^4v_0^3},\nonumber\\
C_A^{(2)}&=&\frac{(2 f-1) \theta _{-1} m^2 v_0^2}{6 v}+\frac{m \left(\theta _1+2 v^3\right)}{3 v^3 v_0},\nonumber\\
C_A^{(5)}&=&\frac{m \left(2 v^3-v_0^2\theta _{-1} \right)}{6 v^2 v_0^2},\qquad\qquad\qquad
C_A^{(6)}~=~\frac{\left(2 v^3-v_0^2\theta_{-1} \right)}{6 v v_0^2}(1-f_p) f_p, 
\end{eqnarray}
with $C_{A,M}^{(3)}=C^{(6)}$, and $C_{A,M}^{(4)}=C^{(7)}$, with $C^{(6,7)}$ defined in (\ref{coe_general}). And the coefficients $C_M^{(i)}$ have the following expression,
\begin{eqnarray}
C_M^{(1)}&=&-\frac{m^2 v_0\theta _{-1} }{3 T^2}(1-f_p) f_p-\frac{ m (2 v^3- v_0^2\theta _{-1})}{6 T v_0^2}(1-2 f_p)-\frac{\theta _1}{3 v_0^3},\nonumber\\
C_M^{(2)}&=&-\frac{m^2 v_0^2\theta_{-1}}{6 T v}(1-2 f_p)+\frac{m (v_0^2\theta_{-1} +v)}{3 v v_0},\nonumber\\
C_M^{(5)}&=&\frac{m v_0^2 (v^3-\theta_1)}{6 v^4},\qquad
C_M^{(6)}~=~\Big(\frac{m\theta_{-1}}{6T}(1-2 f_p)-\frac{2 v^5+\theta _{-1} (1-v^2) v_0^2}{2 v^2 v_0^3}\Big)\frac{(1-f_p) f_p}{2T}. 
\end{eqnarray}
When the spin of the probe reaches global equilibrium, $n^\mu$ takes the solution  (\ref{n_geq}). The global equilibrium expression for $f_A$ and $\mathcal{M}_\perp^\mu$ can be solved accordingly, giving $f_A^{\text{geq}}=-\hat{P}_\perp^\nu\omega_\nu(p/p_0)f_p(1-f_p)/(2T)$ and  $\mathcal{M}_\text{geq}^{\perp,\mu}=\Xi^{\mu\nu}\omega_\nu f_p(1-f_p)/(2T)$. Inserting back into the collision terms in (\ref{fA_M}), one will also find the elimination of collision term at global equilibrium.

In the massless limit, restricted by $\delta(P^2)$, the second and third term in (\ref{decompo_n}) naturally returns to zero. $f_A$ becomes the only degrees of freedom, its transport equation becomes
\begin{eqnarray}
P\cdot \partial f_A=\kappa_{LL}\Big\{\Big(\frac{p(1-f_p)f_p}{3T}+\frac{(1-2f_p)}{3}\Big) f_A
+\frac{p(1-2f_p)}{6}\hat{P}_\perp^\nu \partial_{P_{\perp}^\nu}f_A
-\frac{p T}{6}g^{\nu\rho} \partial_{P_\perp^\nu}\partial_{P_\perp^\rho}f_A
-\frac{f_p(1-f_p)}{6 p}\hat{P}_\perp^\nu\omega_\nu\Big\},
\end{eqnarray}
which is in consistency with \cite{Yang:2020hri}, up to overall constant coefficient.

In the non-relativistic limit, keeping the collision terms to $\mathcal{O}(1/m)$, both transport equations become
\begin{eqnarray}
\label{nonrelativistic}
\Big(\partial_t+\frac{1}{m}P_\perp^\nu\partial_\nu\Big)f_A&=&-\frac{\kappa_{LL}}{9}\Big\{\Big(-\frac{T}{p^2}+\frac{1-2f_p}{m}\Big)f_A+\Big(\frac{5T}{p}-\frac{p(1-2f_p)}{m}\Big)\hat{P}_\perp^\nu \partial_{P_\perp^\nu}f_A+T g^{\nu\rho} \partial_{P_\perp^\nu}\partial_{P_\perp^\rho}f_A+\frac{2T}{m}\partial_{P_\perp^\nu}\mathcal{M}_\perp^\nu\Big\},\nonumber\\
\Big(\partial_t+\frac{1}{m}P_\perp^\nu\partial_\nu\Big)\mathcal{M}_\perp^\mu&=&-\frac{\kappa_{LL}}{9}\Big\{\Big(\frac{5T}{p}-\frac{p(1-2f_p)}{m}\Big)\hat{P}_\perp^\nu \partial_{P_\perp^\nu}\mathcal{M}_\perp^\mu+T\Big(g^{\nu\rho} \partial_{P_\perp^\nu}\partial_{P_\perp^\rho}\mathcal{M}_\perp^\mu-\frac{2}{m v}\hat{P}_\perp^\mu \partial_{P_\perp^\nu}\mathcal{M}_\perp^\nu\Big)+\frac{2T}{m}\Xi^{\mu\nu}\partial_{P_{\perp}^\nu}f_A\Big\}.\nonumber\\
\end{eqnarray}
In the non-relativistic limit, the diffusion of $f_A$ is not suppressed while diffusion of $\mathcal{M}_\perp^\mu$ is suppressed. The coupling between $f_A$ and $\mathcal{M}_\perp^\mu$ is suppressed by $1/m$. The polarization effect for both $f_A$ and $\mathcal{M}_\perp^\mu$ are suppressed by at least $1/m^2$, and thus is excluded in the collision term. 
\subsection{Numerical result}
In order to carry out numerical analysis, we assume for convenience that $f_A$ and $\mathcal{M}_\perp^\mu$ are isotropic in momentum, then both transport equations in (\ref{fA_M}) decouple and can be further simplified. Besides, we ignore the spatial dependence and focus only on the time evolution. We here focus on the evolution of the transverse magnetic dipole moment $\mathcal{M}_\perp^\mu$, through some simple numerical process, its diffusion and polarization can be visualized. The non-linear term $\omega_\perp^\mu$ in the collision term of $\mathcal{M}_\perp^\mu$ indicates that $\mathcal{M}_\perp^\mu$ can get polarized by the transverse components of vorticity. Denoting the direction of $\omega_\perp^\mu$ to be $\hat{x}$, $\mathcal{M}_\perp^\mu$ can be further decomposed into two perpendicular components $\mathcal{M}_{x,y}$. While $\mathcal{M}_{y}$ undergoes a purely diffusion process, $\mathcal{M}_{x}$ gets polarized by the vorticity. Such diffusion and polarization processes also have dependence on mass and momentum. 

\begin{figure}[H]\centering
\includegraphics[height=0.3\textwidth]{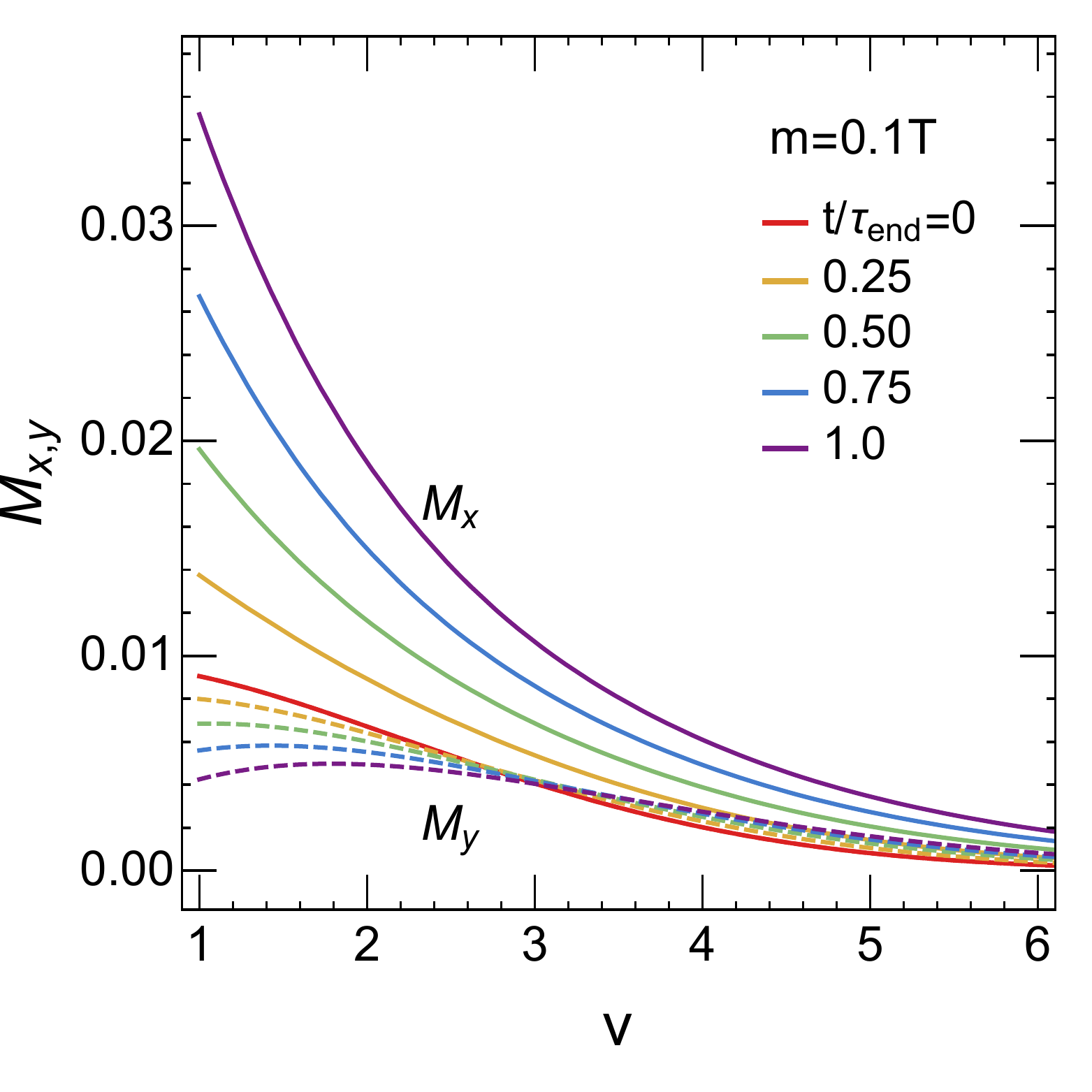}~~
\includegraphics[height=0.3\textwidth]{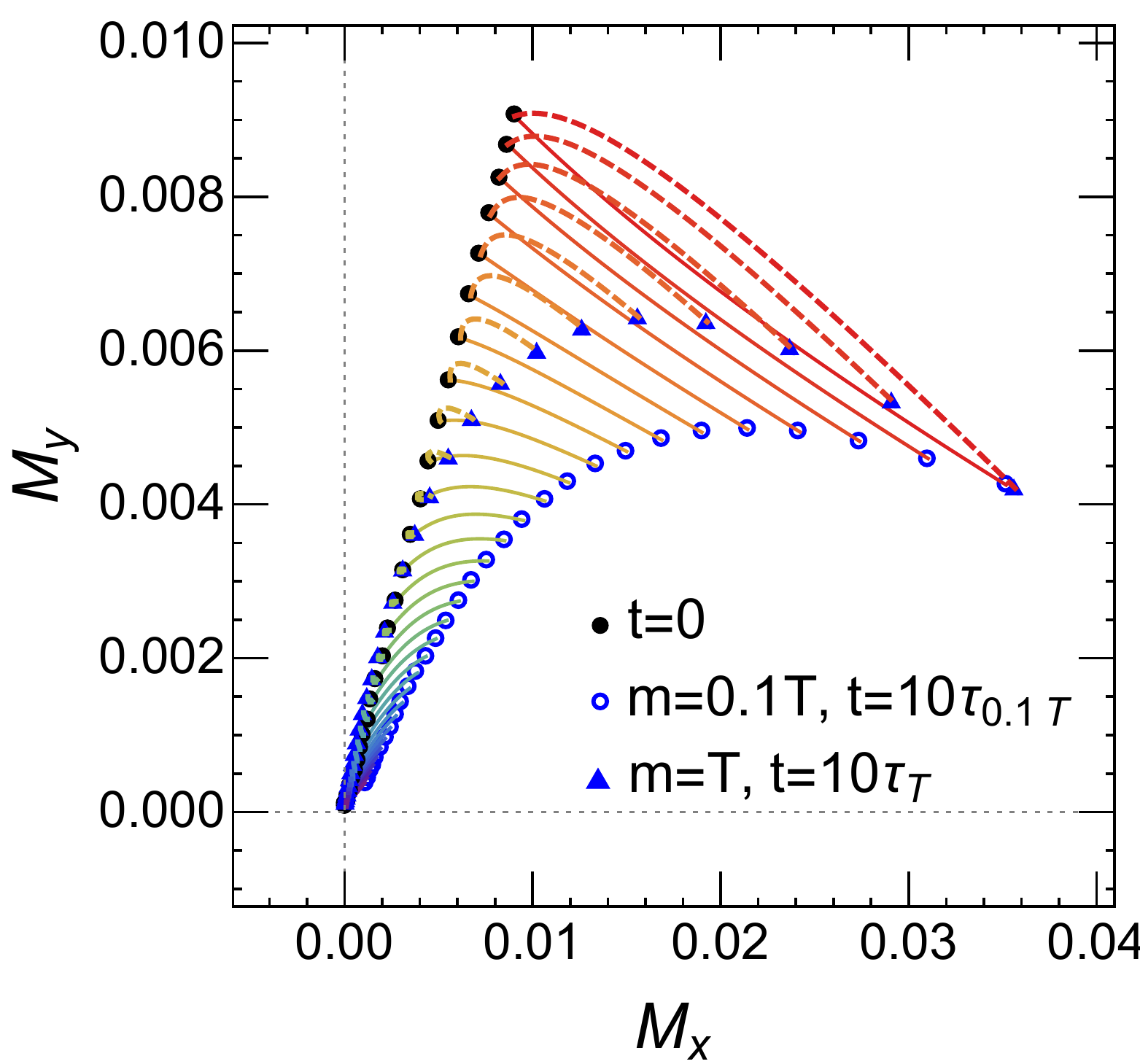}
\caption{Evolution of transverse dipole moment of the probe fermion with initial condition $\mathcal{M}_{x,y}=0.01e^{-v^2/10}$, transverse vorticity is $|\omega_\perp^\mu|=0.2T$. Left: evolution of $\mathcal{M}_{x,y}$ with $m=0.1T$, $\tau_\text{end}=10\tau_{0.1T}$, the solid lines are $\mathcal{M}_{x}$, the dashed lines are $\mathcal{M}_{y}$. Right: compare the evolution of $\mathcal{M}_{x,y}$ with difference mass $m=0.1T$ and $m=T$. The black solid dots are initial condition, the blue circles are $\mathcal{M}_{\perp\mu}$ with mass $m=0.1T$ at $t=10\tau_{0.1T}$, while the blue triangles are $\mathcal{M}_{\perp\mu}$ with mass $m=T$ at $t=10\tau_{T}$. The solid trajectories for $m=0.1T$ and dashed trajectories for $m=T$. Lines from red to purple are low momentum to large momentum. }
\label{fig2}
\end{figure}

We first compare the evolution for difference masses. Taking same gaussian initial condition $\mathcal{M}_{x,y}(t=0,v)=0.01e^{-v^2/10}$, with transverse component of vorticity $|\omega_\perp^\mu|=0.2T$, we compare two different mass of the probe $m=0.1T$ and $m=T$. To guarantee the stability of the evolution, we solve the transport equation from $t=0$ to $t=10\tau_m$, with $\tau_m$ characterizing the relaxation time scale $\tau_m=e^4\ln\frac{1}{e}\frac{T^3}{8\pi^2m^2}$, which depends on mass of the probe. The evolution of transverse dipole moment with $m=0.1T$ is presented in the left panel of Fig.\ref{fig2}, with solid lines denoting $\mathcal{M}_{x}$ and dashed lines for $\mathcal{M}_{y}$. The red line is the initial condition, from red to purple are early to later time in the evolution. One can directly observe that $\mathcal{M}_{x}$ get polarized by $\omega_\perp^\mu$ while $\mathcal{M}_{y}$ experiences some diffusion process. Evolution of the large momentum modes are suppressed compared to low momentum modes. The evolution trajectory of $\mathcal{M}_{\perp\mu}$ with different masses is presented in the right panel of Fig.\ref{fig2}. The black solid dots are initial condition, dots connected with red trajectories are modes with low momentum $v=1$, the trajectories are rainbow colored, with purple lines for modes with large momentum $v=7$. The blue circles are $\mathcal{M}_{\perp\mu}$ with mass $m=0.1T$ at $t=10\tau_{0.1T}$, while the blue triangles are $\mathcal{M}_{\perp\mu}$ with mass $m=T$ at $t=10\tau_{T}$. The solid trajectories for $m=0.1T$ and dashed trajectories for $m=T$. Comparing both sets of solution, one can find that the polarization of $\mathcal{M}_{\perp\mu}$ is strongly suppressed by the mass. This is in consistency with the non-relativistic limit of the transport equation that the polarization effect is at least suppressed by $O(1/m^2)$.
We also solve the transport equations with another set of unphysical but interesting initial conditions, $M_x=0.01e^{-(v-5)^2/5}\cos(\frac{\pi}{8}v)$ and $M_y=0.01e^{-(v-5)^2/5}\sin(\frac{\pi}{8}v)$. Such initial condition gives a circular distribution of initial transverse magnetic dipole moment in the transverse plane. Taking $m=T$, we compare the evolution of $\mathcal{M}_\perp^\mu$ with and without vorticity. In the left panel of Fig.\ref{fig2}, we present the evolution when turning off the transverse vorticity, one can observe a purely diffusion effect. In the right panel, the vorticity is turned on, the low momentum modes experience polarization effect, while the large momentum modes are affected only by diffusion. 
\begin{figure}[H]\centering
\includegraphics[height=0.3\textwidth]{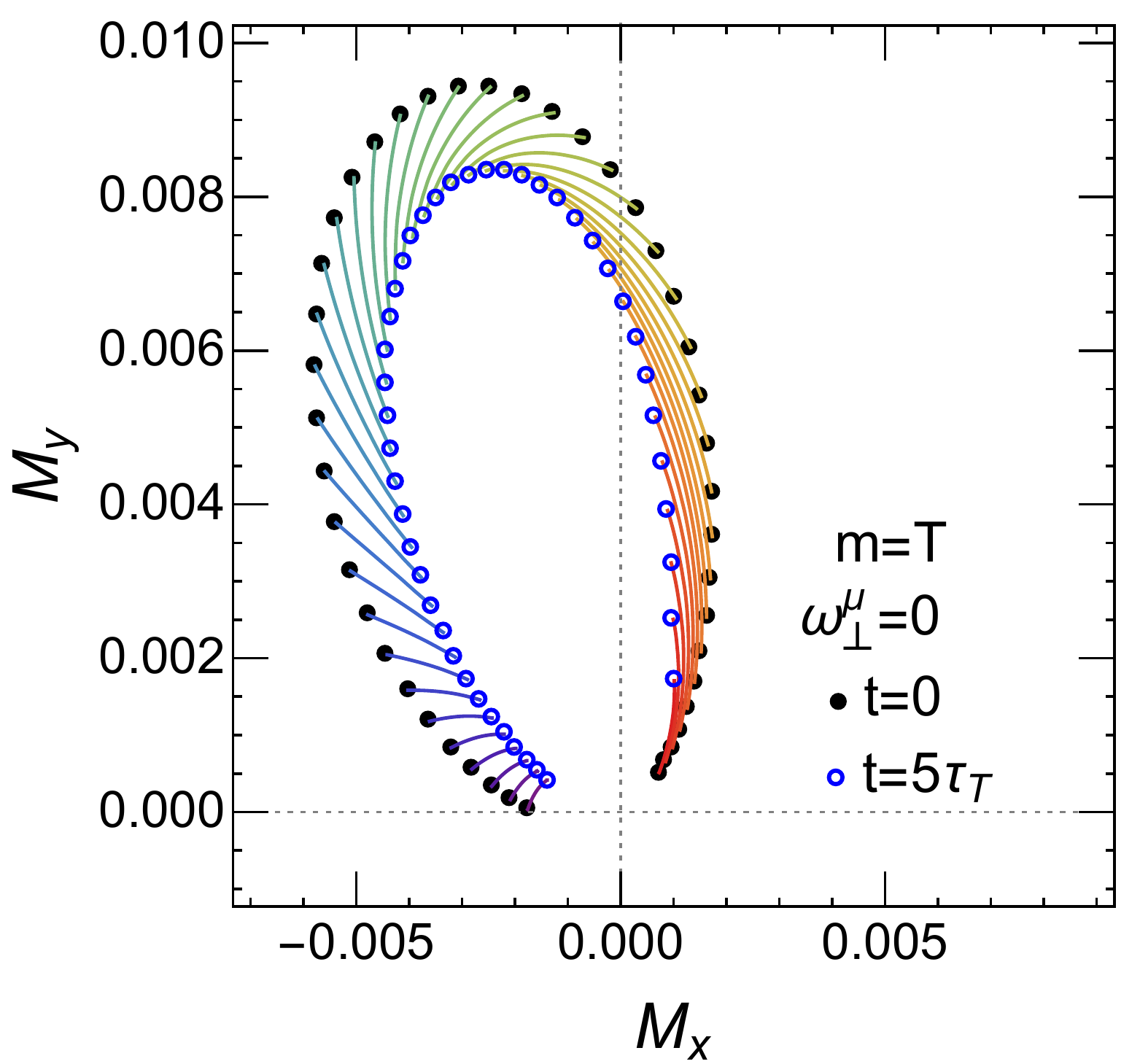}~~
\includegraphics[height=0.3\textwidth]{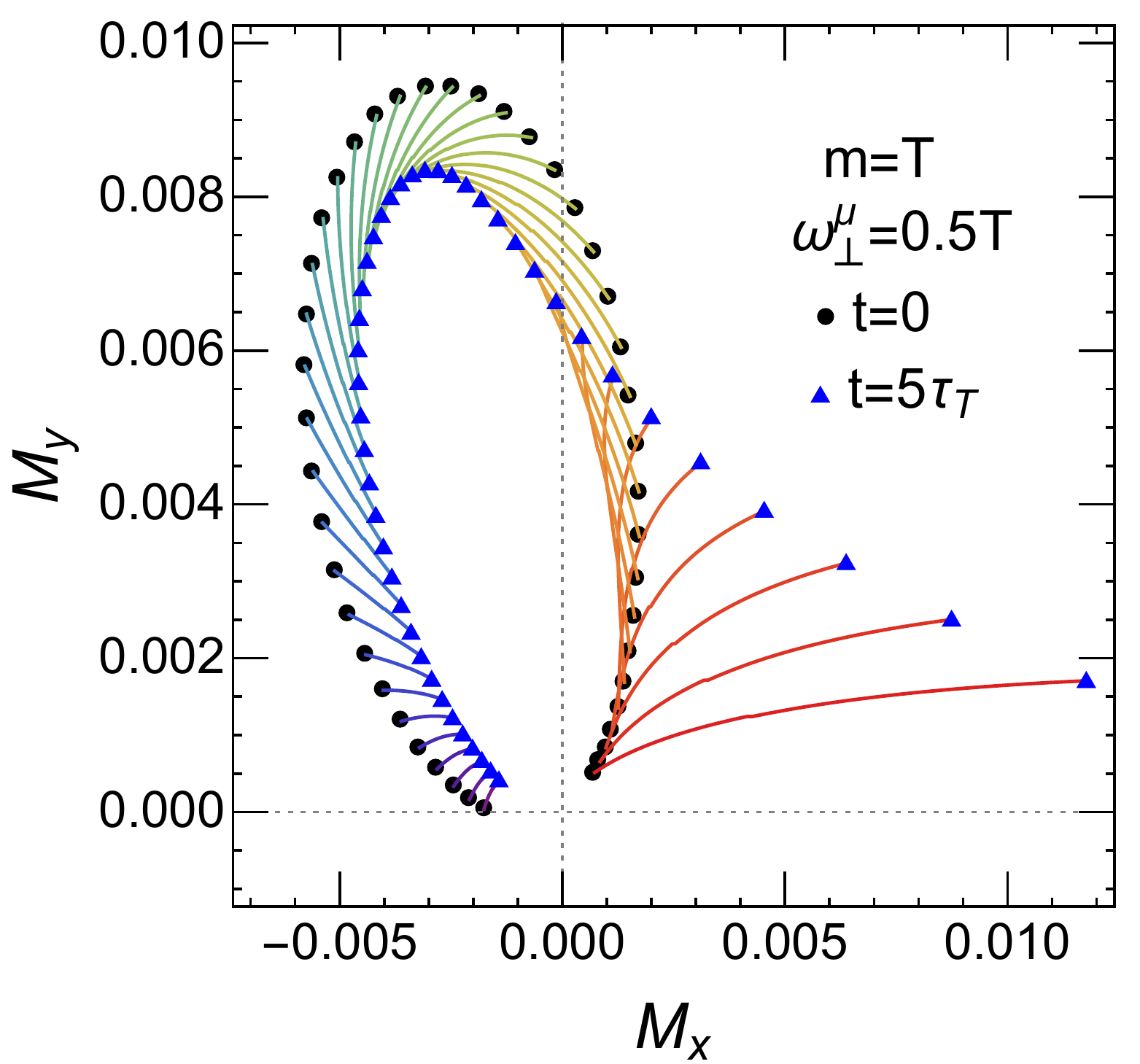}
\caption{Evolution trajectories of transverse dipole moment in the transverse plane, with the vorticity turned off in the left panel, and $|\omega_\perp^\mu|=0.5T$ in the right panel. The red lines for small momentum and purple lines for large momentum.}
\label{fig3}
\end{figure}

Through the above numerical investigation, the diffusion and polarization effect can be obviously observed, as well as the mass dependence. In the numerical solution we have taken the assumption that $\mathcal{M}_\perp^\mu$ is isotropic in momentum space and have ignored the spatial dependence to focus only on the time evolution. The full set of the transport equation can be solved combined with the first order gradients obtained from hydrodynamics, to investigate the spin polarization effect of the massive quarks in the quark-gluon plasma.
\section{Conclusion and Outlook}
\label{s6}
In this paper, we investigate the evolution of spin of a hard massive probe fermion in a massless hot QED plasma at local equilibrium in the framework of quantum kinetic equation. We adopt the assumption that mass of the probe fermion is much larger than the thermal mass $m\gg m_D\sim eT$. In this case, Coulomb scattering contributes to the leading logarithmic collision terms in the HTL approximation, while the Compton scattering is sub-leading and thus excluded. Besides, as the first step, we have assumed that $f_V$ of the probe fermion has reached local equilibrium with the QED plasma, which means only the axial kinetic equation is required. As the axial-charge density and spin polarization in heavy ion collision are aroused by quantum effects or collective motion of the medium characterized by gradients of the fluid velocity, it is physically motivated to count $\mathcal{A}_\mu$ as $\mathcal{O}(\hbar)$. The diffusion and polarization effect in the spin evolution are hence both of $\mathcal{O}(\hbar)$ and should be treated on the same basis. It is thus essential to derive the collision term of the axial kinetic equation to include all first order gradients, or to $\mathcal{O}(\hbar)$ order. 

Under such assumptions, we derive the collision term of axial-kinetic equation to the leading logarithmic order and to the first order of gradients. The diffusion and polarization effects coexist in the collision term, where the former drives the spin fluctuation to damp out, while the later characterizes the spin getting polarized by the vorticity, shear tensor, acceleration and temperature gradients of the fluid. Massless limit and non-relativistic limit of the axial kinetic equation are also presented. The effect of diffusion and polarization balance with each other, leading to the elimination of collision terms in the global equilibrium. Near the global equilibrium, the relaxation rate for the fluctuation is extracted. So as to illustrate the difference among the three dof of $\mathcal{A}_\mu$, the axial kinetic equation is further decomposed into transport equation of axial charge and transverse magnetic dipole moment considering the purely rotational medium. Preliminary numerical analysis is carried out for the transverse dipole moment, the evolution of spin is observed for different mass and magnitude of vorticity. Without vorticity, spin of the probe experiences purely diffusion process, when vorticity is turned on, spin get polarized in direction of the vorticity. Modes with small momentum and small mass get polarized easier, inconsistent with the result of non-relativistic limit. 

The physical settings in this paper can be viewed as a toy model for the evolution of spin of s-quark in the quark gluon plasma. In a more self-consistent scenario, we will consider the scattering of massive quark with a QCD plasma, without assuming the local equilibrium of number distribution for the probe quark, the full set of vector kinetic equation and axial kinetic equation will be derived. Besides, elimination of collision term of axial kinetic equation in global equation is considered in this paper,  local equilibrium for the massive fermion is still under discussion. These will be included in an forthcoming paper. 

\acknowledgments
The author would like to thank Shu Lin and Di-Lun Yang for fruitful discussion and suggestions. The work is supported by NSFC grant Nos. 12005112.

\begin{appendix}
\section{Phase space integral}
\label{calculation_detail}
\subsection{Simplification of integral measure}
\label{integral}
Assuming that the medium fermion and probe fermion are hard fermions with momentum comparable with temperature $p, k, p', k'\sim T$, while the momentum transfer is soft $q_0, q\ll T$, the phase space integral can be simplified using the small momentum transfer as well as the momentum conservation and on-shell condition
\begin{eqnarray}
&&(2\pi)^3\int\frac{d^4Kd^4Qd^4P'd^4K'}{((2\pi)^4)^4}(2\pi)^8\delta(P-K-Q)\delta(Q+P'-K')\epsilon(K\cdot u)\epsilon(P'\cdot u)\epsilon(K'\cdot u)\delta(K^2-m^2)\delta(P'^2)\delta(K'^2)\nonumber\\
&=&\frac{1}{(2\pi)^5}\int dq_0 d^3qd^3k'\frac{1}{2p'_02k'_02k_0}\delta(p_0-k_0-q_0)\delta(p'_0-k'_0+q_0).
\end{eqnarray}
The momentum integral is left with integral over $Q$ and $\vec k'$. It is useful to decompose momentum $\vec q$ and $\vec k'$ into
\begin{eqnarray}
\label{momentum_decomposition}
\vec{k}'&=& k'\cos\theta_k\hat{p}+k'\sin\theta_k\cos\varphi_k\hat{x}+k'\sin\theta_k\sin\varphi_k\hat{y},
\nonumber\\
\vec{q}&=& q\cos\theta_q\hat{p}+q\sin\theta_q\cos\varphi_q\hat{x}+q\sin\theta_q\sin\varphi_q\hat{y},
\end{eqnarray}
where we have denoted $\hat{p}$ as $\hat{z}$ for now. And introduce $\Omega$ as the angle between $\vec k'$ and $\vec q$, namely $\cos\Omega=\cos\theta_k\cos\theta_q-\sin\theta_k\sin\theta_q\cos\Delta\varphi$, with $\Delta\varphi=\varphi_q-\varphi_k$.  The measure can be parametrized as 
\begin{eqnarray}
\int d^3qd^3k'=\int q^2 dq d\cos\theta_q d\varphi_{q} k'^2 dk' d\cos\theta_k d\Delta\varphi.
\end{eqnarray}
Considering that loop fermion are light fermions, which can be treated as massless. 
Using $\vec p'=\vec k'-\vec q$ and $\vec k=\vec p-\vec q$, we can use the on-shell condition to cast the $\delta$-function into 
\begin{eqnarray}
\delta(p_0-k_0-q_0)&\simeq&\delta(q \frac{p}{p_0}\cos\theta_q-q^2\frac{(p^2\sin^2\theta_q+m^2)}{2p_0^3}-q_0),\nonumber\\
\delta(p'_0-k'_0+q_0)&\simeq&\delta(q\cos\Omega-q^2\frac{\sin^2\Omega}{2k'}-q_0),
\end{eqnarray}
with $p_0=(p^2+m^2)^{1/2}$. The angular integral over $\varphi_{q}$ and $\varphi_{k}$ can be performed to obtain, 
\begin{eqnarray}
\int d\varphi_{q}d\varphi_{k}\delta(p'_0-k'_0+q_0)&\simeq&4\pi\frac{1}{q(1+\frac{q_0}{k'})}\frac{1}{[\sin^2\theta_q\sin^2\theta_k-(\cos\Omega-\cos\theta_q\cos\theta_k)^2]^{1/2}}.
\end{eqnarray}
Note that the above $\delta-$function constrain the unique solution of $\cos\Delta\varphi$, yet $\sin\Delta\varphi$ can take both solutions $\pm(1-\cos^2\Delta\varphi)^{1/2}$. Thus integrals containing odd number of $\sin\Delta\varphi$ will be vanishing under the angular integral. The square root constrains the domain of $\cos\theta_k$ as $\cos(\theta_q-\Omega)<\cos\theta_k<\cos(\theta_q+\Omega)$. The other $\delta$-function gives 
\begin{eqnarray}
\int d\cos\theta_q\delta(p_0-k_0-q_0)\simeq\frac{1}{\frac{pq }{p_0}(1+\frac{q_0}{p_0})}.
\end{eqnarray}
From the $\delta$-function, one can solve 
\begin{eqnarray}
&&\cos\Omega\simeq\frac{q_0}{q}+\frac{q}{2k'}\Big(1-\frac{q_0^2}{q^2}\Big)+\mathcal{O}(q^2),\qquad\qquad\quad
\sin\Omega\simeq\Big(1-\frac{q_0^2}{q^2}\Big)^{1/2}\Big(1-\frac{q_0}{2{k'}}\Big),\nonumber\\
&&\cos\theta_q\simeq\frac{p_0q_0}{pq}+\frac{q}{2p}\Big(1-\frac{q_0^2}{q^2}\Big)
+\mathcal{O}(q^2),\qquad\qquad 
\sin\theta_q\simeq\Big(1-\frac{p_0^2}{p^2}\frac{q_0^2}{q^2}\Big)^{1/2}\Big(1-\frac{q^2-q_0^2}{p^2-p_0^2q_0^2}\frac{q_0p_0}{2}\Big),
\end{eqnarray}
in obtaining the leading-log order result, it is enough to keep the above solution to the first order of $q$. Note that $-1\leq \cos\Omega,\cos\theta_q\leq 1$ also set a limit to $x=q_0/q$ that $-\frac{p}{p_0}\leq\frac{q_0}{q}\leq\frac{p}{p_0}$. So that $\int dx dk'$ has the integration domain $\int dx dk'\rightarrow   \int_0^\infty dk'\int_{-p/p_0}^{+p/p_0} dx$. With the above approximation of small momentum transfer, the collision term at leading logarithmic order can be explicitly calculated. The basic process is to collect all terms of integrad to $Q^{-2}$, after combining the measure, and integrate $q$ ranges from $eT\ll q\ll T$, the log thus arises from $\int_{eT}^{T}dq/q=\ln(1/e)$. 

To finish the remaining integral over $k'$, the following expression are often utilized, 
\begin{eqnarray}
\int_0^\infty dk' k' n_F(k')(-1+n_F(k'))&=&-T^2\ln2,\nonumber\\
\int_0^\infty dk' k'^2 n_F(k')(-1+n_F(k'))&=&-\frac{1}{6}\pi^2T^3,\nonumber\\
\int_0^\infty dk' k'^2 n_F^2(k')(-1+n_F(k'))&=&-\frac{1}{6}\pi^2T^3+T^3\ln 2,
\end{eqnarray}
where $n_F(k')=1/(e^{k'/T}+1)$.

\subsection{diffusion}
\label{diffusion_calculation}
In this subsection, we show some details of calculation of diffusion term (\ref{diffusionterm}). The spin diffusion part defined in (\ref{diffusionterm}) is evaluated by first expanding the integrand in terms of $Q$. For this purpose, we use the following expansions 
\begin{eqnarray}
&&M_{A1}^{\mu\nu}n_\nu(P)=+\frac{2}{(Q^2)^2}T_{A1}\,n^\mu(P),\\
&&M_{A2}^{\mu\nu}n_\nu(K)=-\frac{2}{(Q^2)^2}\Big(T_{A2}\,n^\mu(P)+T_{A2,1}^\nu\partial_{P_\perp^\nu} n^\mu(P)+T_{A2,1}^{\nu\rho}\partial_{P_\perp^\rho}\partial_{P_\perp^\nu}n^\mu(P)+T_{A2,2}^{\mu\rho}n_\rho(P)+T_{A2,2}^{\mu\rho\nu}\partial_{P_\perp^\nu}n_\rho(P)\Big),\nonumber
\end{eqnarray}
where the coefficients $T$ in the above expressions are kept to $\mathcal{O}(Q^2)$, giving
\begin{eqnarray}
\label{coeM1M2}
T_{A1}&=&2 (P\cdot K')^2+m^2Q\cdot K'-2Q\cdot K' P\cdot K'-2P\cdot Q P\cdot K'+Q\cdot K' P\cdot Q+Q^2 P\cdot K'+\mathcal{O}(Q^3)\nonumber\\
T_{A2}&=&2 (P\cdot K')^2+m^2Q\cdot K'-2Q\cdot K' P\cdot K'-2P\cdot Q P\cdot K'+Q^2 P\cdot K'+\mathcal{O}(Q^3)\nonumber\\
T_{A2,1}^\nu&=&-Q^\nu(2 (P\cdot K')^2+m^2Q\cdot K'-2Q\cdot K' P\cdot K'-2P\cdot Q P\cdot K') +\mathcal{O}(Q^3)\nonumber\\
T_{A2,1}^{\nu\rho}&=&Q^\nu Q^\rho (P\cdot K')^2  +\mathcal{O}(Q^3)\nonumber\\
T_{A2,2}^{\mu\rho}&=&-2K'^\mu K'^\rho P\cdot Q+2Q^\mu K'^\rho P\cdot K'-P^\mu Q^\rho K'\cdot Q+2K'^\mu Q^\rho(-P\cdot K'+P\cdot Q+Q\cdot K')+\mathcal{O}(Q^3),\nonumber\\
T_{A2,2}^{\mu\rho\nu}&=&2P\cdot K'(K'^\mu Q^\rho Q^\nu-Q^\mu K'^\rho Q^\nu)+\mathcal{O}(Q^3). 
\end{eqnarray}
In obtaining the above expressions, we have used $P^\mu n_\mu(P)=0$ to simplify the derivatives, thus $P^\mu Q^\nu \partial_{P\nu} n_\mu(P)=-Q^\mu n_\mu(P)$, and $P^\mu Q^\rho \partial_{P\rho}Q^\nu \partial_{P\nu} n_\mu(P)=-2Q^\mu Q^\nu \partial_{P\nu} n_\mu(P)$. The leading logarithmic order requires keeping the integrand to $\mathcal{O}(Q^{-2})$, thus it is sufficient to expand the distributions to $\mathcal{O}(Q)$, giving,
\begin{eqnarray}
\bar{f}_K{f}_{P'}\bar{f}_{K'}+{f}_{K}\bar{f}_{P'}{f}_{K'}&=&\bar{f}_{K'}{f}_{K'}-({f}_{P}-\bar{f}_{K'}){f}_{K'}\bar{f}_{K'}\frac{q\cos\Omega}{T}+\mathcal{O}(q^2),\nonumber\\
f_{P}{f}_{P'}\bar{f}_{K'}+\bar{f}_{P}\bar{f}_{P'}{f}_{K'}&=&\bar{f}_{K'}{f}_{K'}-({f}_{K'}-{f}_{P}){f}_{K'}\bar{f}_{K'}\frac{q\cos\Omega}{T}+\mathcal{O}(q^2).
\end{eqnarray}
Taking $T_{A2,2}^{\mu\rho\nu}\partial_{P_\perp^\nu}n_\rho(P)$ for instance to illustrate the integral over such tensor structures. The basic strategy is to convert the integral over the tensor to scalars. After integral, $T_{A2,2}^{\mu\rho\nu}$ will be function of momentum $p$. Besides, as one can obverse, $T_{A2,2}^{\mu\rho\nu}$ is anti-symmetric in exchanging $\mu\rho$, thus can be decomposed into, 
\begin{eqnarray}
\label{TA22}
T_{A2,2}^{\mu\rho\nu}=T_{A2,2}^{(1)}u^{[\mu} g^{\rho]\nu}+T_{A2,2}^{(2)}\hat{P}_\perp^{[\mu} g^{\rho]\nu}+T_{A2,2}^{(3)}\hat{P}_\perp^{[\mu} u^{\rho]}u^\nu+T_{A2,2}^{(4)}u^{[\mu}\hat{P}_\perp^{\rho]}\hat{P}_\perp^{\nu},
\end{eqnarray}
with other projectors vanishing in momentum integral. Using the relations between various projectors, $u_{[\mu} g_{\rho]\nu}T_{A2,2}^{\mu\rho\nu}=6T_{A2,2}^{(1)}-2T_{A2,2}^{(4)}$, $\hat{P}_{\perp[\mu} g_{\rho]\nu}T_{A2,2}^{\mu\rho\nu}=-6T_{A2,2}^{(2)}-2T_{A2,2}^{(3)}$, $\hat{P}_{\perp[\mu} u_{\rho]}u_\nu T_{A2,2}^{\mu\rho\nu}=-2T_{A2,2}^{(2)}-2T_{A2,2}^{(3)}$, $u_{[\mu}\hat{P}_{\perp\rho]}\hat{P}_{\perp\nu}T_{A2,2}^{\mu\rho\nu}=-2T_{A2,2}^{(1)}+2T_{A2,2}^{(4)}$, then each coefficients can be obtained as combinations. The momentum integral of the various scalar functions then follows the processes described in Appendix.\ref{integral}. Giving
\begin{eqnarray}
T_{A2,2}^{(1)}&=&\kappa_{LL}\frac{T}{m v}\frac{m(2v^3-v_0^2\theta_{-1})}{6v^2},\nonumber\\
T_{A2,2}^{(2)}&=&-T_{A2,2}^{(3)}=\kappa_{LL}\frac{T}{m v}\frac{m(2v^3-3v_0^2\theta_{-1})}{6v^2},\nonumber\\
T_{A2,2}^{(4)}&=&-\kappa_{LL}\frac{T}{m v}\frac{mv_0\theta_{-1}}{6v}.
\end{eqnarray}
With the coefficients, the original term becomes 
\begin{eqnarray}
T_{A2,2}^{\mu\rho\nu}\partial_{P_\perp^\nu}n_\rho(P)=T_{A2,2}^{(n)}n_\mu(P)+T_{A2,2}^{(u)}u^\mu +T_{A2,2}^{(p)}\hat{P}_\perp^{\mu}+T_{A2,2}^{(\partial)}\hat{P}_\perp^{\rho}\partial_{P_{\perp\mu}}n_\rho(P),
\end{eqnarray}
where
\begin{eqnarray}
T_{A2,2}^{(n)}&=&\frac{1}{p_0}T_{A2,2}^{(1)},\nonumber\\
T_{A2,2}^{(u)}&=&\frac{p}{p_0}T_{A2,2}^{(1)}\hat{P}_\perp^{\rho}n_\rho(P)+T_{A2,2}^{(1)}\partial_{P_\perp^\rho}n^\rho(P)+T_{A2,2}^{(4)}\hat{P}_\perp^{\rho}\hat{P}_\perp^{\nu}\partial_{P_\perp^\nu}n_\rho(P),\nonumber\\
T_{A2,2}^{(p)}&=&\big(\frac{p^2}{p_0^3}T_{A2,2}^{(1)}+\frac{m^2}{p_0^3}T_{A2,2}^{(4)}\big)\hat{P}_\perp^{\rho}n_\rho(P)+T_{A2,2}^{(2)}\partial_{P_\perp^\rho}n^\rho(P)+\frac{p}{p_0}T_{A2,2}^{(4)}\hat{P}_\perp^{\rho}\hat{P}_\perp^{\nu}\partial_{P_\perp^\nu}n_\rho(P),\nonumber\\
T_{A2,2}^{(\partial)}&=&\frac{p}{p_0}T_{A2,2}^{(1)}-T_{A2,2}^{(2)}.
\end{eqnarray}
Other scalar function or tensors in (\ref{coeM1M2}) are integrated in a similar way. After finishing the detailed calculation, one can arrive at the first two lines in (\ref{general}) and coefficients $C^{(1)}$ to $C^{(7)}$ in (\ref{coe_general}).

\subsection{first oder gradients}
\label{firstorder_calculation}
To calculate the collision terms related to first order gradients including $C_{A\mu}^{\text{vor}}, C_{A\mu}^{\text{shear}}, C_{A\mu}^{\text{Tgra+acc}}$ defined in (\ref{collision-vorticity}), (\ref{collision-shear}) and (\ref{collision_Tgra_acc}). The basic strategy is to converting the momentum integral over tensors into scalars, expanding the integrand to $\mathcal{O}(Q^{-2})$ and taking the momentum integral the same way as in Appendix.\ref{integral}. 
\paragraph{vorticity}
\label{vorticity}: $C_{\mu}^{\text{vor}}$ in the integrand of (\ref{collision-vorticity}) can be further decomposed into
\begin{eqnarray}
C_{\mu}^{\text{vor}}&=&T_\nu^{\text{vor}}\;\omega^{[\mu}u^{\nu]}+T^{\mu\rho\nu}_{\text{vor}}\;\omega_{[\rho}u_{\nu]}.
\end{eqnarray}
The momentum integral over the vorticity part also begins with expansion of the integrand over $Q$, the tensors above are expanded to $\mathcal{O}(Q^{-2})$ giving
\begin{eqnarray}
T_\nu^{\text{vor}}&=&\frac{1}{(Q^2)^2}\Big(2Q_\nu(P\cdot K')^2+(m^2-2P\cdot K')Q\cdot K' Q_\nu+P\cdot Q(Q\cdot K' P_\nu-2P\cdot K'Q_\nu)+\mathcal{O}(Q^3)\Big),\nonumber\\
T^{\mu\rho\nu}_{\text{vor}}&=&
\frac{1}{(Q^2)^2}\Big(2K'^\mu\big(P\cdot Q K'^\rho(P-Q)^\nu+(P\cdot K'-P\cdot Q-K'\cdot Q)Q^\rho P^\nu\big)\nonumber\\
&&\qquad\quad +Q\cdot K' P^{\mu}Q^\rho P^\nu
-2P\cdot K' Q^\mu K'^\rho(P-Q)^\nu+\mathcal{O}(Q^3)\Big),
\end{eqnarray}
since the above two tensors are at least $\mathcal{O}(Q^{-3})$, it is enough to keep $\mathcal{O}(Q)$ order of the distribution functions in order to get the leading logarithmic result,
\begin{eqnarray}
\label{distributionsq}
\bar{f}_{K}\bar{f}_{K'}{f}_{P'}f_{P}=\bar{f}_{K'}{f}_{K'}\bar{f}_{P}{f}_{P}\Big(1+\bar{f}_{K'}\frac{q \cos\Omega}{T}-{f}_{P}\frac{p\,q \cos\theta_q}{p_0 T}\Big)+\mathcal{O}(q^2).
\end{eqnarray}
The momentum integral over tensors $T_\nu^{\text{vor}}$ and $T^{\mu\rho\nu}_{\text{vor}}$ are carried out after transforming the tensors to a series of scalar functions. Since after momentum integral, the vector $T_\nu^{\text{vor}}$ will only be function of momentum $P$, it can be decomposed by 
\begin{eqnarray}
T_\nu^{\text{vor}}&=&u_\nu T_{\text{vor}}^{(1,1)}+\hat{P}_{\perp\nu} T_{\text{vor}}^{(1,2)},
\end{eqnarray}
with $T_{\text{vor}}^{(1,1)}=u^\nu T_\nu^{\text{vor}}$ and $T_{\text{vor}}^{(1,2)}=-\hat{P}_{\perp}^{\nu}T_\nu^{\text{vor}}$. The scalar coefficients can be integrated according to the process in Appendix.\ref{integral}. Then this part becomes $T_\nu^{\text{vor}}\omega^{[\mu}u^{\nu]}=T_{\text{vor}}^{(1,1)}\omega^{\mu}-T_{\text{vor}}^{(1,2)}\hat{P}_{\perp\nu}\omega^{\nu}u^{\mu}$. In the other term $T^{\mu\rho\nu}_{\text{vor}}\omega_{[\rho}u_{\nu]}$, $\omega_{[\rho}u_{\nu]}$ projects out the anti-symmetric part of $T^{\mu\rho\nu}_{\text{vor}}$, thus $T^{\mu\rho\nu}_{\text{vor}}$ can be decomposed similar to (\ref{TA22}), 
\begin{eqnarray}
T^{\mu\rho\nu}_{\text{vor}}=T_{\text{vor}}^{(2,1)}u^{[\nu} g^{\rho]\mu}+T_{\text{vor}}^{(2,2)}\hat{P}_\perp^{[\nu} g^{\rho]\mu}+T_{\text{vor}}^{(2,3)}\hat{P}_\perp^{[\nu} u^{\rho]}u^\mu+T_{\text{vor}}^{(2,4)}u^{[\nu}\hat{P}_\perp^{\rho]}\hat{P}_\perp^{\mu},
\end{eqnarray}
the momentum integral over the various scalar functions $T_{\text{vor}}^{(2,i)}$ can be carried out according to Appendix.\ref{integral}. After obtaining the coefficients, this part will be
\begin{eqnarray}
T^{\mu\rho\nu}_{\text{vor}}\omega_{[\rho}u_{\nu]}=2T_{\text{vor}}^{(2,1)}\omega^\mu-2(T_{\text{vor}}^{(2,2)}+T_{\text{vor}}^{(2,3)})\hat{P}_{\perp\nu}\omega^{\nu}u^\mu +2T_{\text{vor}}^{(2,4)}\hat{P}_{\perp\nu}\omega^{\nu}\hat{P}_\perp^{\mu}.
\end{eqnarray}
Together the above two parts, the vorticity term will be the $C^{(8)}$ term in (\ref{general}), with $C^{(8)}$ defined in (\ref{coe_general}).

\paragraph{shear}
\label{shear}
After momentum integral, the collision term $C_{A\mu}^{\text{shear}}$ (\ref{collision-shear}) will only be function of $P$, thus can in general be expressed in terms of a series of symmetric and traceless projectors as,
\begin{eqnarray}
C_{A\mu}^{\text{shear}}=\left(u_\mu \widehat{Q}_{\alpha\beta}C_{\text{shear}}^{(1)}+P_{\perp\mu}\widehat{Q}_{\alpha\beta}C_{\text{shear}}^{(2)}+\widehat{I}_{\mu\alpha\beta}C_{\text{shear}}^{(3)}+\widehat{T}_{\mu\alpha\beta}C_{\text{shear}}^{(4)}\right)\sigma^{\langle\alpha\beta\rangle}, 
\end{eqnarray}
where the projectors are defined through 
\begin{eqnarray}
\widehat{Q}_{\alpha\beta}&=&\hat{P}_{\perp\alpha}\hat{P}_{\perp\beta}+\frac{1}{3}\Delta_{\alpha\beta},\nonumber\\
\widehat{I}_{\mu\alpha\beta}&=&\hat{P}_{\perp\alpha}\Delta_{\mu\beta}+\hat{P}_{\perp\beta}\Delta_{\mu\alpha}-\frac{2}{3}\hat{P}_{\perp\mu}\Delta_{\alpha\beta},\nonumber\\
\widehat{T}_{\mu\alpha\beta}&=&\frac{1}{2}\left(\epsilon_{\mu\nu\rho\alpha}u^\nu\hat{P}_\perp^\rho \hat{P}_{\perp\beta}+\epsilon_{\mu\nu\rho\beta}u^\nu\hat{P}_\perp^\rho \hat{P}_{\perp\alpha}\right),
\end{eqnarray}
which are symmetric and traceless in the indices $\alpha\beta$. In the local rest frame of the fluid, the above projectors are $Q_{ij}=\hat{p}_i\hat{p}_j-\frac{1}{3}\delta_{ij}$, $I_{kij}=\hat{p}_j\delta_{ik}+\hat{p}_i\delta_{jk}-\frac{2}{3}\hat{p}_k\delta_{ij}$ and $T_{kij}=\frac{1}{2}(\epsilon_{kli}\hat{p}_l\hat{p}_j+\epsilon_{klj}\hat{p}_l\hat{p}_i)$, which are symmetric and traceless in $ij$. Each of the four coefficients $C_{\text{shear}}^{(i)}$ can be obtained by first projecting (\ref{collision-shear}) onto the corresponding projectors and then taking the momentum integral. One will find that only $C_{\text{shear}}^{(4)}$ in non-vanishing. Hence after momentum integral, the shear tensor term in the collision term appears as
\begin{eqnarray}
C_{A\mu}^{\text{shear}}=\widehat{T}_{\mu\alpha\beta}C_{\text{shear}}^{(4)}\sigma^{\langle\alpha\beta\rangle}.
\end{eqnarray}
Using the relation $\widehat{T}^{\mu\alpha\beta}\widehat{T}_{\mu\alpha\beta}=-1$, the evaluating of the contribution from the shear tensor is converted to calculating the scalar function
\begin{eqnarray}
C_{\text{shear}}^{(4)}&=&-4e^4\frac{1}{(2\pi)^5}\int dq_0 d^3qd^3k'\frac{1}{2p'_02k'_02k_0}\delta(p_0-k_0-q_0)\delta(p'_0-k'_0+q_0)\widehat{T}^{\mu\alpha\beta}C_{\mu\alpha\beta}^{\text{shear}}(-\beta)\bar{f}_{K}\bar{f}_{K'}{f}_{P'}f_{P},
\end{eqnarray}
and $C_{\mu\alpha\beta}^{\text{shear}}$ is defined in (\ref{C-shear}), which can be simplified into, 
\begin{eqnarray}
C_{\mu\alpha\beta}^{\text{shear}}&=&+\epsilon_{\mu\alpha\sigma\lambda}\Big(-\frac{P'\cdot K'}{2}K^\sigma P^\lambda K_\beta+K\cdot K' P'^\sigma P^\lambda P_\beta-P\cdot P' K'^\sigma K^\lambda K_\beta+\frac{m^2 Q\cdot K'}{P'\cdot u}P'^\sigma u^\lambda P'_\beta\Big)\nonumber\\
&&-\epsilon_{\xi\alpha\sigma\lambda}\Big(P^\sigma K^\lambda P'^\xi K_\beta+\frac{m^2}{P'\cdot u}Q^\xi P'^\sigma u^\lambda P'_\beta
\Big)K'_\mu +\{P'\leftrightarrow K'\},
\end{eqnarray}
where $\{P'\leftrightarrow K'\}$ part is to taking conversion accordingly in the above all terms. Using 
\begin{eqnarray}
\label{two_epsilon}
\epsilon^{\mu\nu\rho\sigma}\epsilon_{\mu\nu\alpha\beta}&=&-2\delta_{\alpha\beta}^{\rho\sigma}=-2(\delta_\alpha^\rho\delta_\beta^\sigma-\delta_\alpha^\sigma\delta_\beta^\rho),\nonumber\\
\epsilon^{\mu\nu\rho\sigma}\epsilon_{\mu\xi\alpha\beta}&=&-\delta_{\xi\alpha\beta}^{\nu\rho\sigma}=-[\delta_\xi^\nu(\delta_\alpha^\rho\delta_\beta^\sigma-\delta_\alpha^\sigma\delta_\beta^\rho)-\delta_\alpha^\nu(\delta_\xi^\rho\delta_\beta^\sigma-\delta_\xi^\sigma\delta_\beta^\rho)
+\delta_\beta^\nu(\delta_\xi^\rho\delta_\alpha^\sigma-\delta_\xi^\sigma\delta_\alpha^\rho)], 
\end{eqnarray}
to finish the contractions in $\widehat{T}^{\mu\alpha\beta}C_{\mu\alpha\beta}$.  Since $\widehat{T}^{\mu\alpha\beta}C_{\mu\alpha\beta}$ is at least $\mathcal{O}(Q)$, it is enough to expand the distribution functions to $\mathcal{O}(Q)$ (\ref{distributionsq}). After taking the momentum integral according to Appendix.\ref{integral}, one will obtain the $C^{(9)}$ term in (\ref{general}), with $C^{(9)}$ in (\ref{coe_general}).

\paragraph{temperature gradients and acceleration}
\label{Tandacc}
Both of the tensors $C_{\mu\lambda}^{\text{Tgra}}$ and $C_{\mu\lambda}^{\text{acc}}$ defined in (\ref{C_Tgra_acc}) can be expanded in a general structure, namely
\begin{eqnarray}
T_{\mu\lambda}(P)=T_1 g_{\mu\lambda}+T_2 u_{\mu}u_\lambda+T_3 \hat{P}_{\perp\mu}\hat{P}_{\perp\lambda}+T_4u_{\mu}\hat{P}_{\perp\lambda}+T_5\hat{P}_{\perp\mu}u_\lambda+T_6\epsilon_{\mu\lambda\alpha\beta}u^\alpha\hat{P}_\perp^\beta, 
\end{eqnarray}
where one can find only projector $\epsilon_{\mu\lambda\alpha\beta}u^\alpha\hat{P}_\perp^\beta$ have non-vanishing coefficient under the momentum integral. Using the relation $\epsilon^{\mu\lambda\rho\sigma}u_\rho\hat{P}_{\perp\sigma}\epsilon_{\mu\lambda\alpha\beta}u^\alpha\hat{P}_\perp^\beta=2$, (\ref{collision_Tgra_acc}) can be casted into 
\begin{eqnarray}
C_{A\mu}^{\text{Tgra+acc}}&=&
C_{\text{Tgra}}\epsilon_{\mu\lambda\alpha\beta}u^\alpha\hat{P}_\perp^\beta\partial^\lambda\ln T+ C_{\text{acc}}\epsilon_{\mu\lambda\alpha\beta}u^\alpha\hat{P}_\perp^\beta D u^\lambda, 
\end{eqnarray}
with 
\begin{eqnarray}
C_{\text{Tgra}}&=&-\frac{4e^4}{(2\pi)^5}\int dq_0 d^3qd^3k'\frac{1}{2p'_02k'_02k_0}\delta(p_0-k_0-q_0)\delta(p'_0-k'_0+q_0)\frac{1}{2}\epsilon^{\mu\lambda\rho\sigma}u_\rho\hat{P}_{\perp\sigma}C_{\mu\lambda}^{\text{Tgra}}(-\beta)\bar{f}_{K}\bar{f}_{K'}{f}_{P'}f_{P},\nonumber\\
C_{\text{acc}}&=&-\frac{4e^4}{(2\pi)^5}\int dq_0 d^3qd^3k'\frac{1}{2p'_02k'_02k_0}\delta(p_0-k_0-q_0)\delta(p'_0-k'_0+q_0)\frac{1}{2}\epsilon^{\mu\lambda\rho\sigma}u_\rho\hat{P}_{\perp\sigma}C_{\mu\lambda}^{\text{acc}}(-\beta)\bar{f}_{K}\bar{f}_{K'}{f}_{P'}f_{P}.
\end{eqnarray}
Using (\ref{two_epsilon}) to complete the contraction, and carrying out the momentum integral according to Appendix.\ref{integral}, one will obtain the $C^{(10)}$ and $C^{(11)}$ term in (\ref{general}), with $C^{(10)}$ and $C^{(11)}$ in (\ref{coe_general}).

\section{gauge issue}
\label{gaugeissue}
One can explicitly check that the collision term is gauge independent. In this section, we show for example that terms related to vorticity is gauge independent. Photon propagator in temporal axial gauge, Coulomb gauge and covariant gauge are given by
\begin{eqnarray}
\label{gauges}
\text{temporal axial gauge:}\quad G_{\mu\nu}&=&\frac{-1}{Q^2}P^T_{\mu\nu}+\frac{-1}{Q^2}\Big(\frac{Q^2}{q^2}u_\mu u_\nu-\frac{Q^2}{q_0 q^2}u_{(\mu}Q_{\nu)}+\frac{Q^2}{q_0^2 q^2}Q_\mu Q_\nu\Big),\nonumber\\
\text{Coulomb gauge:}\quad G_{\mu\nu}&=&\frac{-1}{Q^2}P^T_{\mu\nu}+\frac{-1}{Q^2}\;\frac{Q^2}{q^2}u_\mu u_\nu,\nonumber\\
\text{covariant gauge:}\quad G_{\mu\nu}&=&\frac{-1}{Q^2}P^T_{\mu\nu}+\frac{-1}{Q^2}\Big(\frac{Q^2}{q^2}u_\mu u_\nu-\frac{q_0}{q^2}u_{(\mu}Q_{\nu)}+\frac{q^2}{q_0^2 Q^2}Q_\mu Q_\nu\Big).
\end{eqnarray}
The above covariant gauge corresponds to the Landau gauge $\xi=0$
\begin{eqnarray}
G_{\mu\nu}=\frac{1}{Q^2} \Big(g_{\mu\nu}-(1-\xi)\frac{Q_\mu Q_\nu}{Q^2}\Big).
\end{eqnarray}
while in the calculation we have adopted Feynman gauge $\xi=1$. The point is to work out the one-loop corrected photon propagator $G_{\mu\nu}^{(0,1)}$ in various gauges and to check whether the different terms among various gauges are vanishing under momentum integral. The expression of $G_{\mu\nu}^{(0,1)}$ in Landau gauge is given by (\ref{photon_propa}). Feynman gauge and Landau gauge differs only in tensor structure of $Q_\mu Q_\nu$. While the three gauges in (\ref{gauges}) differs by $Q_\mu Q_\nu$, $u_{(\mu}Q_{\nu)}$ and crossing terms with $P^T_{\mu\nu}$ when multiplying two photon propagators. 

To check the gauge invariance of zeroth order photon propagator $G^{(0)<}_{\mu\nu}(Q)=D_{\mu\beta}^{22}(Q)D_{\alpha\nu}^{11}(Q)\Pi^{(0)<\alpha\beta}(Q)$, one can find that contracting $\Pi^{(0)<\alpha\beta}$ with $Q_\alpha Q_\beta Q_\mu Q_\nu$ and $g_{\mu\alpha}Q_\beta Q_\nu+g_{\nu\beta}Q_\mu Q_\alpha$ will both leads to vanishing results under the $\delta$-function. Thus Feynman gauge and Landau gauge give the same $G^{<(0)}_{\mu\nu}$. Temporal axial gauge and Coulomb gauge can be shown to give also the same $G^{<(0)}_{\mu\nu}$. 

We then check the vorticity terms in first order photon propagator $G^{(1)<}_{\mu\nu}(Q)=D_{\mu\beta}^{22}(Q)D_{\alpha\nu}^{11}(Q)\Pi^{(1)<\alpha\beta}(Q)$. The vorticity related terms in $\Pi^{(1)<\alpha\beta}(Q)$ is
\begin{eqnarray}
\Pi^{(1)<\alpha\beta}_\omega(Q)=-2i\epsilon^{\alpha\rho\beta\sigma}\int_{P',K'}K'_\sigma (P'\cdot \omega u_\rho-P'\cdot u\omega_\rho)\bar{f}_{K'}f'_{P'}-P'_\rho (K'\cdot \omega u_\sigma-K'\cdot u\omega_\sigma)f_{P'}f'_{K'}.
\end{eqnarray}
As only self-energy components $\Sigma_{A\mu}$ and $\Sigma_{T\mu\nu}$ contains the first order photon propagator. The collision term involving $G^{(1)<}_{\mu\nu}(Q)$ in (\ref{Atransport}) can be extracted out using the expression of the components of self-energy (\ref{selfenergy_components}), giving,
\begin{eqnarray}
\label{G1incollision}
&&\int_{K,Q}i\Big((m^2-P\cdot K)\epsilon_{\mu\nu\rho\sigma}Q^\nu
- P_\mu\epsilon_{\lambda\nu\rho\sigma}Q^\nu P^{\lambda}\Big) \Big(G^{(1)<\mu\nu}(Q)f_P\bar{f}_K-G^{(1)>\mu\nu}(Q)\bar{f}_P {f}_K \Big)\nonumber\\
&=&\int_{K,Q}i\Big((m^2-P\cdot K)\epsilon_{\mu\nu\rho\sigma}Q^\nu
- P_\mu\epsilon_{\lambda\nu\rho\sigma}Q^\nu P^{\lambda}\Big) G^{\mu\alpha}_R G^{\nu\beta}_A\Big(\Pi^{(1)<}_{\alpha\beta}(Q)f_P\bar{f}_K-\Pi^{(1)>}_{\alpha\beta}(Q)\bar{f}_P {f}_K \Big).
\end{eqnarray}
Although the expression of $G^{(1)<}_{\mu\nu}(Q)$ depends on gauge choice, one can explicit show that the collision term involving $G^{(1)<}_{\mu\nu}(Q)$ (\ref{G1incollision}) is not gauge depending. This can be proved by contracting 
\begin{eqnarray}
&&\Pi^{(1)<\alpha\beta}_\omega f_P\bar{f}_K -\Pi^{(1)>\alpha\beta}_\omega \bar{f}_P{f}_K\nonumber\\
&\propto&\frac{1}{2T}\epsilon^{\alpha\beta\rho\sigma}\Big(
Q\cdot\omega K'_\rho u_\sigma
-K'\cdot\omega Q_\rho u_\sigma
+K'\cdot u Q_\rho\omega_\sigma
-Q\cdot u K'_\rho \omega_\sigma
\Big)\bar{f}_{K'}\bar{f}_Kf_P{f}_{P'}
\end{eqnarray}
with $Q_\alpha Q_\beta Q_\mu Q_\nu$, $g_{\mu\alpha}Q_\beta Q_\nu+g_{\nu\beta}Q_\mu Q_\alpha$, $u_{(\alpha} Q_{\mu)}u_{(\beta} Q_{\nu)}$ and crossing terms respectively, then further projecting to $((m^2-P\cdot K)\epsilon_{\mu\nu\rho\sigma}Q^\nu- P_\mu\epsilon_{\lambda\nu\rho\sigma}Q^\nu P^{\lambda})$. One will find that it will either be directly vanishing by symmetry, or be proportional to $\vec{q}\times\vec{k}'\cdot\vec{\omega}$ which is vanishing under momentum integral. In this way, the vorticity related term is also gauge independent. Other terms related to first order gradient are also proved to be gauge independent in a similar way.

\end{appendix}

\end{document}